\documentclass[11pt]{article}

\usepackage{graphicx} 
\usepackage{amsfonts}
\usepackage{amsmath}
\usepackage{amssymb}
\usepackage{amsthm}
\usepackage[mathscr]{eucal}
\usepackage{bbold}
\usepackage{dsfont}
\usepackage{framed}
\usepackage{jheppub}
\usepackage{makecell}  
\usepackage{mathtools}
\usepackage{multicol}
\usepackage{physics}
\usepackage[normalem]{ulem}
\usepackage{soul}  
\usepackage{tensor}
\usepackage{tikz}
\usetikzlibrary{math} 
\usepackage{thmtools}
\usepackage{thm-restate}
\usepackage{xcolor}
\usepackage{cleveref}
\usepackage{colortbl}
\usepackage{enumitem}
\usepackage{subcaption}

\newtheorem{conjecture}{Conjecture}

\newtheorem{defi}{Definition}
\newtheorem{lemma}{Lemma}

\newtheorem{thm}{Theorem}

\newcommand{\shadeTF}{\cellcolor{red!5}}
\newcommand{\shadeG}{\cellcolor{gray!5}}

\newcommand{\shadeI}{\cellcolor{blue!5}}
\newcommand{\shadeR}{\cellcolor{red!5}}

\newcommand{\shadeO}{\cellcolor{orange!5}}
\newcommand{\shadeT}{\cellcolor{teal!5}}

\newcommand{\hlorange}[1]{\sethlcolor{orange!25}\hl{#1}}  
\newcommand{\hllime}[1]{\sethlcolor{lime!25}\hl{#1}}      
\newcommand{\hlcyan}[1]{\sethlcolor{cyan!25}\hl{#1}}      

\newcommand{\Ink}[1]{\color{brown}{\{#1\}}}
\newcommand{\Cnk}[1]{\color{violet}{\{#1\}}}
\newcommand{\ino}{$i\#$}
\newcommand{\cno}{$c\#$}

	\newcommand{\ctform}{tripartite form}
	
	\newcommand{\qtor}{toric}
		\newcommand{\qcyc}{dihedral}
	\newcommand{\qproj}{projective}

	\newcommand{\nN}{{\sf N}}  
	\newcommand{\nD}{{\sf D}}  

\definecolor{Qncolor}{rgb}{0,0,0}
\definecolor{QTcolor}{rgb}{0,0,0}
\definecolor{QPcolor}{rgb}{0,0,0}

	\newcommand{\Qn}[1]{{\color{Qncolor}{\tensor*{Q}{^{[#1]}}}}}   
	\newcommand{\QT}[1]{{\color{QTcolor}{\tensor*[^{^\text{T}\!}]{Q}{_{(#1)}}}}}   
	\newcommand{\QP}[1]{{\color{QPcolor}{\tensor*[^{^\text{P}\!}]{Q}{_{(#1)}}}}}   

	\newcommand{\pI}{{\mathscr{I}}}
	
	\newcommand{\pK}{{\mathscr{K}}}

	\newcommand{\x}[1]{\text{\textnormal{\small{#1}}}}  
        \newcommand{\xs}[2]{
            \x{#1}_{#2}
        } 
        \newcommand{\xm}[3]{
            \x{#1}_{#2}^{(#3)}
        } 
        
        \newcommand{\xtp}[2]{
            \x{#1}_{#2}^{+}
        } 
        \newcommand{\xtm}[2]{
            \x{#1}_{#2}^{-}
        } 
        \newcommand{\xx}[2]{#1${}_{#2}$} 
        
	\newcommand{\xS}[1]{S_{_{\text{{#1}}}}}  
	\newcommand{\xI}[1]{{\tt{I}}_{_{\text{{#1}}}}}  

	\newcommand{\sI}{{\tt{I}}}  

	\newcommand{\xtI}[1]{I_3({\text{{\small{#1}}})}}  
	\newcommand{\xctI}[2]{I_3({\text{{\small{#1}}}}|{\text{{\small{#2}}}})}  

	\newcommand{\tI}[1]{I_3({\text{{#1}})}}  
	\newcommand{\ctI}[2]{I_3({\text{{#1}}}|{\text{{#2}}})}  

    \newcommand{\fS}[1]{S\!\left( #1 \right)}
	\newcommand{\ftI}[3]{I_3 \! \left( #1 \, \text{\textnormal{:}} \, #2 \, \text{\textnormal{:}} \, #3 \right)}  
	\newcommand{\fctI}[4]{I_3 \! \left( #1\, \text{\textnormal{:}} \, #2 \, \text{\textnormal{:}} \, #3 \, | \, #4 \right)}  

	\newcommand{\seq}[1]{$#1$}

\renewcommand{\arraystretch}{1.35}
\input{TFmacro.tex} 

\title{Tripartite Form Universality in Holographic Entropy Inequalities}

\author[a]{Veronika E. Hubeny}
\emailAdd{veronika@physics.ucdavis.edu}
\author[b,c]{and Yu Liu}
\emailAdd{yu.liu@epfl.ch}
 
\affiliation[a]{Center for Quantum Mathematics and Physics (QMAP)\\ 
Department of Physics \& Astronomy, University of California, Davis, CA 95616, USA}
\affiliation[b]{Department of Physics, Tsinghua University, Beijing 100084, China}
\affiliation[c]{Institute of Physics, Ecole Polytechnique Federale de Lausanne (EPFL), CH-1015 Lausanne, Switzerland}

\abstract{
To elucidate the meaning of holographic entropy inequalities (beyond subadditivity) which characterize the entanglement structure of geometric states in holography, \cite{Hernandez-Cuenca:2023iqh} proposed the ``tripartite form'' for these inequalities, consisting of tripartite information and conditional tripartite information terms with unit coefficients.  While this provides a compact and useful packaging of the inequalities, it is not a priori guaranteed that all inequalities can be recast in this form.  Here we conjecture that they can, and present substantial evidence, by proving that the two known infinite families of holographic entropy inequalities found in \cite{Czech:2023xed} can indeed be written in the tripartite form.  This is significant because such recasting is  particularly nontrivial for these families.
Apart from providing the explicit tripartite form expressions for every member of these two infinite families, we detail how we arrived at them, in the process deriving several useful identities which may serve as stepping stones to formulate further repackaging of the corresponding information quantities.
We also illustrate the power of the tripartite form by proving a number of structural properties satisfied by any holographic entropy inequality.
}

\begin{document}
\maketitle

\section{Introduction}
\label{s:intro}

One of the long-standing goals of holographic dualities has been to understand how the bulk spacetime emerges from the dual boundary description.  A promising avenue to elucidate this mechanism considers the entanglement structure of  \emph{geometric}\footnote{\, 
    These refer to states in holographic CFTs which encode a classical bulk geometry.
    We allow any spacetime dimensionality and time dependence compatible with the usual physical requirements; we assume the spacetime to be asymptotically AdS, obey Einstein's equation, and satisfy the null energy condition.  (However, the explicit criteria are tangential to the present work; they merely underlie the expectation that the inequalities discussed below pertain to any geometric state rather than solely a static one, as recently emphasized in \cite{Grado-White:2025jci,Grimaldi:2025jad}.)
} states in holography.
It is then natural to focus on the relations between entanglement entropies of spatially-delimited subsystems on the boundary, and extract the key ingredients which underlie the geometric nature of the state.
Subsystem entropies for arbitrary quantum states are known to obey two fundamental universal inequalities, subadditivity (SA) and strong subadditivity (SSA), which can be respectively expressed as the positivity of mutual information and of conditional mutual information.
Geometric states additionally satisfy more stringent relations, known as \emph{holographic entropy inequalities} (HEIs),\footnote{\, 
    By HEI we mean a \emph{primitive} \cite{Hubeny:2018trv} HEI, one which cannot be obtained as a conical combination of other HEIs and is therefore non-redundant.  In the geometric language of the holographic entropy cone of \cite{Bao:2015bfa}, such an HEI is associated with a facet of the cone.
    \label{fn:primitive}
} the number of which depends on the number $\nN$ of elementary subsystems (or ``parties'') specified.  

For $\nN=3$, the only new inequality is the monogamy of mutual information (MMI) \cite{Hayden:2011ag}, or equivalently the negativity of tripartite information (which in turn renders SSA redundant).  However as $\nN$ increases, these inequalities become both more numerous and more intricate.  The full structure is known only up to $\nN=5$ \cite{Bao:2015bfa,Cuenca:2019uzx}, where there are 372 inequalities, organized into 8 orbits under permutation and purification symmetry, some involving over 20 entropies.   At $\nN=6$, only a (likely very small) subset of the inequalities has been found \cite{Hernandez-Cuenca:2023iqh}, which already has 1877 distinct orbits (comprising 8,665,853 distinct inequalities, some involving as many as 30 entropies).\footnote{\, 
    Subsequently, additional $\nN=6$ HEIs have been found (and the list is still growing, cf.\ \cite{hecdata}), but in this paper only use the HEIs from \cite{Hernandez-Cuenca:2023iqh}.
}

Unlike SA and MMI, these higher HEIs have no simple interpretation.  Written in the so-called S-basis (i.e.\ directly in terms of subsystem entanglement entropies), there is no discernible pattern.  It is therefore desirable to uncover a more compact packaging of these expressions. The hopeful expectation is that once one can identify the optimal building blocks, the operational meaning of these quantities will become apparent.  This motivated \cite{He:2019ttu} to explore the I-basis (consisting of $n$-party multipartite informations between individual parties \cite{Hubeny:2018ijt}) and the K-basis (formulated in terms of perfect tensor structures \cite{Cui:2018dyq}).  However, while these reduced the number of terms in the $\nN=5$ HEIs, the simplification didn't suffice to reveal the underlying structure.  

A more substantial improvement was attained in \cite{Hernandez-Cuenca:2023iqh}, which introduced the \emph{tripartite form} (TF) of the HEIs.  This consists of the tripartite informations and conditional tripartite informations, with fixed unit coefficients but arbitrary composite arguments, explained in more detail in \cref{ss:ctform}; cf.~\cref{eq:ctform}.
Unlike the S-, I-, and K- bases representations of the HEIs, the TF rendition is typically not unique, in the sense that a given expression can be rearranged in different combinations of (conditional) tripartite informations.  However, any TF expression automatically implements a number of desirable properties, listed in  \cref{ss:ctform}, which are required for any higher HEI.  The most prominent property is superbalance, defined in \cite{Hubeny:2018ijt} and argued in \cite{He:2020xuo} to be a property of all HEIs (other than SA which is special and which we henceforth exclude from our discussion).  
However, it is not the case that any TF expression need be sign-definite; a simple example is any single conditional tripartite information \cite{Hernandez-Cuenca:2023iqh}: for such an information quantity one can  find geometric states for which it is negative as well as those for which it is positive.

More interestingly, while not all information quantities can be written in a TF, based on the observation that all the $\nN\le5$ HEIs can, \cite{Hernandez-Cuenca:2023iqh} used the TF ansatz to find over 1800 new HEI orbits, and conveniently explore some of their properties.  In particular, the authors proved that none of the corresponding information quantities can constitute correlation measures, because superbalance precludes monotonicity under partial tracing.  
More recently, the utility of the TF was demonstrated in \cite{Grimaldi:2025jad}, where it was used to  characterize what happens to HEIs (written in TF) under ``null reductions'' and in particular to prove that all such null reductions give valid (albeit redundant) inequalities, which then provides a novel characterization of the holographic entropy cone.\footnote{\, 
    An alternate proof which didn't rely on TF was subsequently presented in \cite{Grimaldi:2026lbq}, but this work likewise made use of TF, specifically for conveniently generating false superbalanced inequalities. 
}
Yet another potential use of the TF relates to the tantalizing structural similarities between distinct HEIs exhibited by the TF for $\nN \le 6$ HEIs  \cite{Hernandez-Cuenca:2023iqh}.  These suggest that one might use the TF for generating higher $\nN$ HEIs more systematically, with the ultimate goal of characterizing the full holographic entropy cone for any $\nN$.\footnote{\, 
    A separate route to this goal proceeds by characterizing the cone by its extreme rays rather than its facets; the program initiated in \cite{Hernandez-Cuenca:2019jpv,Hernandez-Cuenca:2022pst} relates these to the extreme rays of a more primitive construct, the subadditivity cone (explored recently for $\nN=6$ in \cite{He:2022bmi,He:2023cco,He:2023aif,He:2024xzq,SACwip} and repackaged in terms of more general structure in \cite{Hubeny:2024fjn}; see also \cite{Hubeny:2025bjo,Hubeny:2025hst}); it would be very interesting to relate these two (dual) programs to each other. 
}

However, this hope, as well as all the above-indicated utility, is predicated on all HEIs being writable in a tripartite form in the first place, or said more concisely, being \emph{TF-compatible}.  The fact that all \emph{known} $\nN \le 6$ HEIs are TF-compatible is encouraging but it is only a rather weak evidence since most of them are so by construction: \cite{Hernandez-Cuenca:2023iqh} scanned over all TF expressions of a given length and determined which ones constitute a valid HEI.  Nevertheless, given the success of finding so many HEIs, we formulate this as a conjecture (also stated recently in \cite[Conj.5]{Grimaldi:2025jad}):
\begin{conjecture}
\label{conj:HEIsTF}
    Every information quantity corresponding to a superbalanced HEI is TF-compatible, meaning it can be written in a \ctform, \cref{eq:ctform}.
\end{conjecture}

To test this conjecture, we need an independently-generated set of inequalities.  Just such a fortuitous set is provided by the two infinite families of HEIs (formulated for arbitrary odd values of $\nN$) introduced in \cite{Czech:2023xed} and proved to be primitive in \cite{Czech:2024rco}.  
These are the so-called ``toric'' and ``projective plane'' (or $\mathbb{RP}^2$) inequalities, reviewed in \cref{ss:inffam}; here we denote the corresponding information quantities by $\QT{m,n}$ (for the toric ones) and $\QP{m}$ (for the projective ones), where $m$ and $n$ are integers which label the individual members of these families.  
The toric family subsumes a previously-proposed infinite family \cite{Bao:2015bfa} known as ``cyclic'' (or more correctly ``dihedral'') inequalities; the 5-party \qcyc\ inequality being perhaps the most familiar higher HEI beyond MMI.
Both of these generalize MMI by keeping distinct discrete symmetries, and can be expressed surprisingly compactly in the S-basis.  For this very reason, they constitute an ideal testing ground for \cref{conj:HEIsTF}: because they are so compact in the S-basis, they are in fact much lengthier when re-cast in the I-basis, which is however the more natural starting point for extracting a tripartite form.  So in a sense, these inequalities should be the hardest to rewrite in a TF.  
Showing that one can nevertheless do so  therefore constitutes a highly nontrivial test and strong evidence for \cref{conj:HEIsTF}. 

In this paper, we find that \emph{both} infinite families of HEIs are indeed TF-compatible, and we provide explicit expressions for their tripartite forms, cf.~\cref{eq:torct} and \cref{eq:projct}.  In fact, these forms are gratifyingly compact in the TF.  For example, $\QP{7}$ has 97 terms in the S-basis, 1023 terms in the I-basis, and only 21 terms in the TF.  More generally, at large index the I-basis expressions contain exponentially many terms, while the S-basis and the TF expressions scale only quadratically; see \cref{tab:scaling}.
Ironically, rewriting the \qcyc\ family turns out to be the most non-trivial step in the whole construction.
In uncovering this structure, one also finds several useful relations between the different variants of TF, and likewise structural similarities between the two families. 

The organization of this paper is as follows.  In \cref{s:background} we briefly review the necessary background material.  After setting up the context of entropy inequalities, we review the tripartite form in \cref{ss:ctform} (where we also introduce a new diagrammatic representation of this form), and the two infinite families of HEIs in \cref{ss:inffam}.  
We then proceed to derive the TF expressions for both families, the \qtor\ one in \cref{s:toric} and the \qproj\ one in \cref{s:proj}.  Since the primary aim is to motivate and explain the construction, to avoid breaking the flow we relegate the longer proofs of the main lemmas and theorems to later sections.  \Cref{s:building} focuses on the behavior of the two families under purifications, which utilizes a new structural identity and serves to highlight the connections between  the \qcyc\ and \qproj\ HEIs.  The remaining proofs are collected in appendix \ref{app:proof}, which also presents alternate forms of the \qcyc\ family.
Finally, \cref{s:discussion}  concludes with a discussion of the TF merits for the two families, and, in order to illustrate the power of \cref{conj:HEIsTF}, highlights a number interesting structural properties of general HEIs which follow directly from TF-compatibility.

\section{Background}
\label{s:background}

We will mostly follow the conventions and notation of \cite{Hernandez-Cuenca:2023iqh} and of \cite{Czech:2023xed} as convenient.  For $\nN$ elementary subsystems $\x{A}_i$ with $i\in [\nN] \equiv \{1,2,\ldots,\nN\}$ (also referred to as ``parties'', corresponding to specified boundary spatial regions), we can form $\nD=2^{\nN}-1$ independent composite subsystems $X_\pI$ with $\pI \subseteq [\nN]$, and associate entanglement entropy $S(X_\pI) \eqcolon S_\pI$ to each of them.  We can then express an \emph{information quantity} $Q$ in the S-basis as a linear combination of these entanglement entropies,
\begin{equation}\label{eq:SbasisIQ}
    Q = \sum_{\pI \subseteq [\nN]} q_\pI \, S_\pI \ .
\end{equation}
A \emph{holographic entropy inequality} (HEI) can be written in the form $Q \ge 0$, with specific coefficients $q_\pI$ in \cref{eq:SbasisIQ};  WLOG we can take $q_\pI \in \mathbb{Z}$ \cite{Bao:2015bfa}.  The actual value of $Q$ depends on the state and region specification, but the positivity for an HEI must hold for \emph{any} geometric state and region specification.

\paragraph{I-basis:}
A special family of information quantities are the multipartite informations $\sI_n$ (in the information theory literature known as interaction information), with its arguments given by $n$ of the elementary subsystems $\x{A}_i$, such as $\sI_n(\x{A}_1 : \x{A}_2 : \ldots : \x{A}_n)$, for which we adopt the shorthand $\sI_\pI$ with $\pI$ comprising of all the arguments and $n=\abs{\pI}$ left implicit.
These can be constructed from the subsystem entropies with unit coefficients and sign which only depends on the size of the subsystem, 
\begin{equation}\label{eq:IfromS}
    \sI_\pI = \sum_{\pK \subseteq \pI} (-1)^{\abs{\pK}+1} \, S_\pK \ ,
\end{equation}
and the full collection of all $\nD$ of them constitutes another basis, the so-called I-basis \cite{Hubeny:2018ijt,He:2019ttu}.  Any information quantity $Q$ can be equivalently written in the I-basis as
\begin{equation}\label{eq:IbasisIQ}
    Q = \sum_{\pI \subseteq [\nN]} r_\pI \, \sI_\pI \ ,
\end{equation}
where the coefficients $r_\pI$ can be obtained from \cref{eq:SbasisIQ} using the expression for individual entropies in I-basis,
\begin{equation}\label{eq:SfromI}
    S_\pI = \sum_{\pK \subseteq \pI} (-1)^{\abs{\pK}+1} \, \sI_\pK \ .
\end{equation}

The I-basis has several convenient features, absent from the S-basis. For any valid HEI $Q\ge 0$, the sign of the coefficients $r_\pI$ is determined purely from the parity of $\abs{\pI}$; in particular each $\sI_n$ comes with $(-1)^n$ sign \cite{Hubeny:2018ijt,Hernandez-Cuenca:2023iqh}.  This poses a restriction on the allowed form of any HEI.
The I-basis also makes it immediate to check if a given information quantity $Q$ is \emph{superbalanced} \cite{Hubeny:2018ijt}, defined by $r_\pI = 0 \ \forall \ \abs{\pI} \le 2$ (a property which must likewise hold for all higher HEIs \cite{He:2020xuo}).  In other words, $Q$ is superbalanced if its I-basis expression only involves tripartite and higher multipartite informations.\footnote{\, \label{fn:Rbalance} 
	More generally, we say $Q$ is \emph{$R$-balanced} if its I-basis form has $r_\pI = 0 \ \forall \ \abs{\pI} \le R$ \cite{Hubeny:2018ijt}.  (Hence 1-balanced is synonymous with balanced, and 2-balanced with superbalanced.)  It is an empirical observation that no known HEI is $R$-balanced for $R\ge 3$, and a theorem that all HEIs with the exception of SA must be superbalanced \cite{He:2020xuo}.
}  In the S-basis, superbalance corresponds to the property that for each singleton $\x{A}_i$ and for each doubleton $\x{A}_i\x{A}_j$, the sum of all coefficients containing these terms vanishes,
\begin{equation}\label{eq:superbalanceS}
    \sum_{\pI \ni i}  q_\pI = 0 \ \ \forall \ i
    \qq{and}
    \sum_{\pI \supseteq ij}  q_\pI = 0 \ \ \forall \ i,j  \ ,
\end{equation}
the former condition defining a balanced quantity.

As pointed out in \cite{Hernandez-Cuenca:2023iqh}, the I-basis provides a natural  characterization of a given information quantity, and in particular the number of terms of various sizes indicate the level of its complexity.  This was formalized in \cite{Hernandez-Cuenca:2023iqh} as the \emph{\ino\ classification}, consisting of an ordered tuple of numbers corresponding to the number of all $\sI_n$ terms of size $n$ (when written with unit coefficients in \cref{eq:IbasisIQ}).  For superbalanced information quantities, we can truncate this to 
$\Ink{i_3,i_4,i_5, \ldots}$ (since $i_1=i_2=0$), where $i_n = \# \sI_n$.\footnote{\, 
	For TF-compatible $Q$, we can also drop all the vanishing terms at the higher $n$ end since once $i_k=0$ for some $k>3$, then $i_n=0$ for all $n>k$ as well.
}
Since each information quantity $Q$ has a unique \ino\ classification associated to it, we can use it to guess at a possible rewriting of $Q$ in terms of  more convenient building blocks (such as the TF, introduced more formally in \cref{ss:ctform}), by identifying the \ino\ classification for each building block and making sure they sum up to the requisite \ino\ classification for $Q$.

\subsection{Tripartite form representation}
\label{ss:ctform}

As indicated in the Introduction, the I-basis representation of an HEI does not constitute a sufficient enough improvement to the S-basis representation, so we seek a more natural set of building blocks.  Since all HEIs (except SA which we are ignoring) are superbalanced, we don't need to invoke entropies and mutual informations, but only start with tripartite information.  On the other hand, any higher multipartite information $\sI_n$ with $n>3$ can be rewritten in terms of just tripartite information with composite arguments, by iterating identities of the type\footnote{\, 
	Notice the change in font, following \cite{Hernandez-Cuenca:2023iqh}, from multipartite information $\sI_n$ with single-party arguments (comprising the basis elements of the I-basis) to the more general multipartite information $I_n$ with optionally composite arguments.
}
\begin{equation}\label{eq:I4fromI3}
	I_{n+1}(\x{A:B:C:D:}\ldots) = I_{n}(\x{A:C:D:}\ldots) + I_{n}(\x{B:C:D:}\ldots) - I_{n}(\x{AB:C:D:}\ldots) \ .
\end{equation}
So we could recast any superbalanced information quantity purely as a linear combination of tripartite information terms ($n=3$), but this now produces an expression where not all $I_3$'s come with the same sign.  However, a  closer observation reveals that for the HEIs in the I-basis representation, each $\sI_4$ term comes accompanied with a $-\sI_3$ term with arguments contained in the $\sI_4$ arguments, leaving behind an expression which can be re-expressed as a conditional tripartite information,
\begin{equation}\label{eq:CI3fromI3}
	-\xctI{B:C:D}{A} = \xtI{A:C:D} - \xtI{AB:C:D} \ ,
\end{equation}
and an analogous version of this simplifying behavior seems to hold for the higher $I_n$ terms as well.
This empirical observation motivated \cite{Hernandez-Cuenca:2023iqh} to guess that for HEIs, a positive sum of negative tripartite informations and negative conditional tripartite informations would suffice to recast the full expression.  

More formally, we define the \ctform\ as follows (cf.\ \cite[Defn.2]{Hernandez-Cuenca:2023iqh}):
\begin{defi}[TF]\label{def:ctform}
An information quantity $Q$ is said to be in the \emph{\ctform}, or equivalently to be \emph{TF-compatible}, if it can be expressed as
\emph{
\begin{equation}\label{eq:ctform}
    Q = \sum_i - I_3(\x{X}_i : \x{Y}_i : \x{Z}_i \, | \, \x{W}_i )
\end{equation}
} with a finite number of tripartite information (when \emph{$\x{W}_i=\emptyset$}) and conditional tripartite information terms, with arguments corresponding to disjoint subsystems 
\emph{$\x{X}_i \cup \x{Y}_i \cup \x{Z}_i \cup \x{W}_i \subseteq [\nN]$}.
\end{defi}

Notice that all coefficients are fixed to be $-1$, but by repeating terms with the same arguments we can capture any negative integer.  As can be easily verified, this form automatically guarantees superbalance as well as  the $\sI_n$ sign alternation mentioned above.  On the other hand, it does not guarantee positivity, nor does it guarantee another crucial property of HEIs, namely that a ``purification'' (discussed below) of a TF expression remains TF-compatible. 

However, another key feature of a TF is that for a given TF-compatible information quantity (consisting of more than just pure $\sI_3$ terms), there are multiple ways of rewriting the expression.  This is already evident from \cref{eq:CI3fromI3}, since each term manifests different permutation symmetries.\footnote{\, 
	The fact that not all terms of the form $\xctI{X:Y:Z}{W}$ are independent of each other immediately follows from the fact that there are many more of them than the dimensionality $\nD$ of the \emph{entropy space}, where each information quantity is associated with a hyperplane (or equivalently covector).  In fact, already the number of (unconditioned) tripartite informations is given by the Stirling number of the second kind 
	 $ \{\smqty{\nN+2 \\ 4} \}$ which scales  with $\nN$ as $\sim \frac{2}{3}4^\nN$ which exponentially faster than $\nD\sim 2^\nN$.
}  For example, by implementing the $\{\x{A,B}\}$ permutation symmetry in the last term of \cref{eq:CI3fromI3}, we can swap which party gets conditioned on:
\begin{equation}\label{eq:multTFs}
	- \xtI{AB:C:D} 
	\ =\  - \xctI{B:C:D}{A} - \xtI{A:C:D} 
	\ =\  - \xctI{A:C:D}{B} - \xtI{B:C:D} \ .
\end{equation}
Each of the three equivalent expressions in \cref{eq:multTFs} constitutes a valid TF.
This freedom, as well as strategies for finding a TF starting from $Q$ rendered in the I-basis, were explained in further detail in \cite[App.B]{Hernandez-Cuenca:2023iqh}.

\paragraph{TF \ino\ classification:}
Despite the multitude of different tripartite form expressions into which one can generically recast a given TF-compatible information quantity $Q$, an invariant characterization of its complexity is the  \ino\ classification.  It will therefore be useful to tabulate the \ino\ classification for the various types of (conditional) tripartite informations.  Since the \ino's are invariant under permuting the party labels, the distinct possibilities are characterized solely by the cardinalities of the arguments (and in fact further invariant under permuting the first three arguments).
Let us schematically denote by $I_{(i:j:k)}$ the tripartite information with argument sizes $\{i,j,k\}$ (for example $\xtI{A:B:CD}$ would be represented by $I_{(1:1:2)}$), and by $C_{(i:j:k|\ell)}$ the conditional tripartite information with argument sizes $\{i,j,k,\ell\}$ with the last corresponding to the conditioned-on subsystem (for example $\xctI{A:B:CD}{EFG}$ is represented by $C_{(1:1:2|3)}$).  If we iteratively apply relations of the type \cref{eq:multTFs}, noting that starting from any such identity we can obtain a new one by conditioning each term on the (same) new party, we can relate the \ino\ classification of each $C_{(1:1:k|\ell)}$ to the Pascal's triangle, 

{\scriptsize
\begin{equation}\label{eq:Pascal}
{\setlength{\arraycolsep}{1pt}
\renewcommand{\arraystretch}{0.75}
\begin{array}{ccccccccc}
&&&&1&&&&\\
&&&1&&1&&&\\
&&1&&2&&1&&\\
&1&&3&&3&&1&\\
1&&4&&6&&4&&1 \\
&&&.&.&.&&&\
\end{array}}
\end{equation}
}
For example, the \ino\ for $C_{(1:1:1|\ell)}$ is given by the $(\ell+1)^{\text{st}}$ row (with the first column corresponding to $\#\sI_3$), and the \ino\ for $I_{(1:1:k)}$ is given by removing the first entry from  the $(k+1)^{\text{st}}$ row.  Moreover, each time we condition an expression on another party, the \ino\ gets augmented by the same \ino\ shifted over to the right by 1.
We apply these observations below (cf.~\cref{tab:inc} and \cref{tab:sing}) to motivate a guess for rewriting the HEIs of interest in a TF.

\paragraph{STF and \cno\ classification:}
While the original motivation for TF was its compactness, we can vastly reduce the redundancy at the expense of compactness, by writing it in a reduced form involving just $C_{(1:1:1|\ell)}$ for various $\ell$, in other words where in each term the three principal (unconditioned) arguments are all singletons.  Accordingly, we dub this the \emph{singleton-tripartite form} (STF).  Any TF-compatible expression can be easily converted into STF by iterating \cref{eq:multTFs} to expand out each term with composite unconditioned arguments.
For example,\footnote{\, 
    A more general formula for $I_{(1:1:k)}$ with arbitrary $k$ will be presented in \cref{eq:TFtoSTFgen} using a more convenient notation.
}
\begin{equation}\label{eq:toSFT}
  \xtI{A:B:CDE} = \xctI{A:B:C}{DE} + \xctI{A:B:D}{E} + \xtI{A:B:E} \ ,
\end{equation}
and the same identity holds with overall conditioning, such as
\begin{equation}\label{eq:toSFTcond}
  \xctI{A:B:CDE}{F} = \xctI{A:B:C}{DEF} + \xctI{A:B:D}{EF} + \xctI{A:B:E}{F} \ .
\end{equation}

In \cref{eq:toSFT} and \cref{eq:toSFTcond} the LHS expression is in TF but not STF, whereas the RHS is in STF.  
While this simplification still doesn't fix the redundancy completely  (as evidenced already by \cref{eq:multTFs}), it does have the key advantage that the number of terms of $C_{(1:1:1|\ell)}$ type for a given $\ell$ is uniquely fixed.  In particular, a single $C_{(i:j:k|\ell)}$ term gets expanded into $ijk$ STF terms of the form $C_{(1:1:1|q)}$ with $q$ ranging from $\ell$ to $\ell+i+j+k-3$.  

The above observation then provides an alternate indicator of the complexity of a TF-compatible information quantity $Q$: instead of the \ino\ classification  $\Ink{i_3,i_4,i_5,\ldots}$, we can use the \cno\ classification, written as $\Cnk{c_0,c_1,c_2,\ldots}$, where we denote by $c_\ell$ the number of $C_{(1:1:1|\ell)}$ terms in STF, with $c_0$ representing the number of unconditioned singleton tripartite information terms.
We can easily convert between these two classifications, and indeed both are related to the Pascal's triangle (in particular they can be expressed simply in terms of binomial coefficients).
Specifically, the number $i_n$ of $\sI_n$ singleton multipartite terms appearing in $C_{(1:1:1|\ell)}$ is $i_n = {\ell \choose {n-3}}$,
and we can invert this to write the $i_n$ coefficient in $c_\ell$ as $(-1)^{n+\ell+1} \, {n-3\choose \ell}$, where this is defined to vanish when $\ell>n-3$ or equivalently $n<\ell+3$.
For example, given the \ino\ classification $\Ink{i_3,i_4,i_5,i_6,i_7,i_8}$, we need $c_k$ only up to $k=5$, with:
\begin{equation}\label{eq:i2c}
\small
\begin{split}
    c_0 &=i_3-i_4+i_5-i_6+i_7-i_8 \\
    c_1 &= i_4 - 2 \,  i_5 + 3  \, i_6 - 4  \, i_7 + 5  \, i_8 \\
    c_2 &= i_5 - 3  \, i_6 + 6  \, i_7 - 10  \, i_8 \\
    c_3 &= i_6 - 4 \,  i_7 + 10 \,  i_8 \\
    c_4 &= i_7 - 5 \,  i_8 \\
    c_5 &= i_8
\end{split}
\end{equation}
Notice the alternating sign in \cref{eq:i2c}; this makes the \cno\ classification more packageable since it consists of smaller numbers.  Furthermore, the non-negativity of $c_\ell$'s poses a restriction on the $i_n$ distribution, which provides a necessary condition for TF-compatibility.  (Similarly, non-negativity of $i_n$'s places a restriction on $Q$ which provides a necessary condition for $Q$ to be an HEI.)

\paragraph{Diagrammatic representation:}
To visualize a given TF, and correspondingly the relations between different TF expressions, we invent the following compact diagrammatic representation:  Each TF of the form \cref{eq:ctform} is represented by table, with rows representing the individual (conditional tripartite information) terms and columns to individual parties $\x{A}_j$.   The colors in the corresponding boxes encode the arguments.  More specifically, the $i$th row corresponds to the term $- I_3(\x{X}_i : \x{Y}_i : \x{Z}_i \, | \, \x{W}_i )$, and the box color in the $j$th column is red if $\x{A}_j \subseteq \x{X}_i$, green  if $\x{A}_j \subseteq \x{Y}_i$, blue  if $\x{A}_j \subseteq \x{Z}_i$, black  if $\x{A}_j \subseteq \x{W}_i$, and white  if $\x{A}_j \nsubseteq \x{X}_i \cup \x{Y}_i \cup \x{Z}_i \cup \x{W}_i $ (in other words, the $ij$th box is blank if $\x{A}_j$ does not appear in any of the arguments of the $i$th term).

Each information quantity expressed in a given TF is then represented by one such table; for example $-\xctI{A:B:C}{D}$ at $\nN=4$ is represented by 
\vTFtab[np]{4}{{A,B,C,D}}.
Similarly, a relation between multiple information quantities can be presented as a relation between the tables.  For example, at $\nN=4$, \cref{eq:multTFs} would be presented as:
\begin{center}
    \vTFtab[np]{4}{{AB,C,D}}
    \ = \ \vTFtab[np]{4}{{B,C,D,A},{A,C,D}}
    \ = \ \vTFtab[np]{4}{{A,C,D,B},{B,C,D}} 
\end{center}
Using the diagrammatic representation has the advantage that it is easier to perceive the pattern and develop a more intuitive feel for it.  For instance, to understand how the above factorization can be iterated in various ways, consider $-\xtI{ABC:DE:F} $ at $\nN=6$:
\begin{center} 
    \vTFtab[np]{6}{{ABC,DE,F}}
    \ = \ \vTFtab[np]{6}{{A,DE,F},{B,DE,F,A},{C,DE,F,AB}}
    \ = \ \vTFtab[np]{6}{{ABC,D,F},{ABC,E,F,D}}
    \ = \ \vTFtab[np]{6}{{A,D,F},{B,D,F,A},{C,D,F,AB},{A,E,F,D},{B,E,F,AD},{C,E,F,ABD}}
    \ = \ \vTFtab[np]{6}{{A,D,F},{A,E,F,D},{B,D,F,A},{B,E,F,AD},{C,D,F,AB},{C,E,F,ABD}} 
\end{center}
Notice that the order of the rows is immaterial: it corresponds to the order of summing the terms in \cref{eq:ctform}.  (The last two tables, reduced to STF, are in fact just row-permutations of each other.)  Furthermore, in each row, we can permute the red, green, and blue colors.  On the other hand the black color boxes, indicating the conditioned-on arguments, are not interchangeable with the others.  

\paragraph{Examples of $\nN \le 6$ HEIs:}
One of the main results of \cite{Hernandez-Cuenca:2023iqh} is that \emph{all known $\nN\le 6$ HEIs are TF-compatible.}
To appreciate the simplification of the expression, in \cref{tab:HEI_N6} we present a few examples of $\nN\le 6$ HEIs from \cite{Hernandez-Cuenca:2023iqh}, where we denote the corresponding information quantity by $\Qn{n}$, with $n$ corresponding to the identifier in the ancillary files of \cite{Hernandez-Cuenca:2023iqh}, also available from \cite{hecdata} (though for ease of comparison, we sometimes chose some other instance in the given permutation orbit).  For each representative  $\Qn{n}$, we provide three alternative expressions, first in the S-basis (white background), second in the I-basis (blue background), and third in a TF (pink background).  In order to fit the former two expressions compactly, we adopt the shorthand notation used in \cite{Hernandez-Cuenca:2023iqh}, where entropies of subsystems are indicated by a subscript e.g.\ $S(\x{ABCD}) \eqcolon \xS{ABCD}$ and likewise the singleton $\sI_n$ arguments are indicated by a subscript, e.g.\ $\sI_4(\x{A:B:C:D}) \eqcolon \xI{ABCD}$.  


%
\begin{table}[htbp] 
\begin{center}
\scriptsize
\begin{tabular}{| c | c | }
\hline
label &  $\Qn{n}$ HEIs for $\nN\le 6$ \\ 
\hline
\hline
$\Qn{2}$
    &  \makecell{$  -\xS{ABC}-\xS{A}-\xS{B}-\xS{C}  $
        \\$ +\, \xS{AB}+\xS{AC}+\xS{BC} $}
\\ 
    & \shadeI{$ -\xI{ABC} $} 
\\ 
   & \shadeTF{$ -\tI{A:B:C} $}
\\  \hline 
$\Qn{5}$
    &  \makecell{$  -\xS{ABCD}-\xS{ACDE}-\xS{AB}-\xS{AD}-\xS{DE}-\xS{C}  $
        \\$ +\, \xS{ABC}+\xS{ABD}+\xS{ACD}+\xS{ADE}+\xS{CDE} $}
\\ 
    & \shadeI{$ \xI{ABCD}+\xI{ACDE}-\xI{ACD}-\xI{ACE}-\xI{BCD} $} 
\\ 
   & \shadeTF{$ -\tI{AB:C:D}-\ctI{A:C:E}{D} $}
\\  \hline 
$\Qn{9}$  
    &  \makecell{$  -\xS{ABCD}-\xS{ACDE}-\xS{BCDE}-\xS{AB}-\xS{AD}-\xS{BC}-\xS{CD}-\xS{CE}-\xS{DE}  $
        \\$ +\, \xS{ABC}+\xS{ABD}+\xS{ACD}+\xS{ADE}+\xS{BCD}+\xS{BCE}+2\xS{CDE} $}
\\ 
    & \shadeI{$ \xI{ABCD}+\xI{ACDE}+\xI{BCDE}-\xI{ACD}-\xI{ACE}-\xI{BCD}-\xI{BDE} $} 
\\ 
  & \shadeTF{$ -\tI{AB:C:D}-\ctI{B:D:E}{C}-\ctI{A:C:E}{D} $}
\\  \hline 
$\Qn{10}$  
    &  \makecell{$  -\xS{ABCE}-\xS{ACDE}-\xS{BCDE}-\xS{ABD}-\xS{AC}-\xS{AE}-\xS{BC}-\xS{BE}-\xS{CD}-\xS{DE}  $
        \\$ +\, \xS{ABC}+\xS{ABE}+\xS{ACD}+\xS{ACE}+\xS{ADE}+\xS{BCD}+\xS{BCE}+\xS{BDE}+\xS{CDE} $}
\\ 
    & \shadeI{$ \xI{ABCE}+\xI{ACDE}+\xI{BCDE}-\xI{ABD}-\xI{ACE}-\xI{BCE}-\xI{CDE} $} 
\\ 
     & \shadeTF{$ -\tI{A:B:D}-\ctI{B:C:E}{A}-\ctI{C:D:E}{B}-\ctI{A:C:E}{D} $}
\\  \hline 
$\Qn{23}$ 
    &  \makecell{$  -\xS{ABCDE}-\xS{ABCDF}-\xS{ABDEF}-\xS{CDF}-\xS{AD}-\xS{AF}-\xS{BC}-\xS{BE}-\xS{BF}-\xS{DE}-\xS{A}  $
        \\$ +\, \xS{ABDE}+\xS{ACDF}+\xS{BCDE}+\xS{BCDF}+\xS{ABC}+\xS{ABF}+\xS{ADE}+\xS{ADF}+\xS{BEF} $}
\\ 
    & \shadeI{$ -\xI{ABCDE}-\xI{ABCDF}-\xI{ABDEF}+2\xI{ABCD}+\xI{ABCE}+\xI{ABCF}+\xI{ABDE}+2\xI{ABDF}+\xI{ABEF}+\xI{ACDE}$} 
\\  & \shadeI{$+\xI{ADEF}+\xI{BDEF} -\xI{ABC}-2\xI{ABD}-\xI{ABE}-\xI{ABF}-\xI{ACD}-\xI{ACE}-\xI{AEF}-\xI{BDF}-\xI{DEF} $} 
\\ 
 & \shadeTF{$ -\tI{A:BC:DE}-\tI{AD:BE:F}-\ctI{A:B:CD}{F} $}
\\ 
 \hline 
\end{tabular}
\end{center}
\caption{Few of the HEI information quantities for $\nN \le 6$ in the S-basis (white background), I-basis (blue background), and the \ctform\ (pink background). We label the information quantities in the form $\Qn{n}$, with $n$ corresponding to the identifier in the ancillary files of \cite{Hernandez-Cuenca:2023iqh}.
The notational shorthand is such that e.g.\ $\xS{ABCD} \coloneqq S(\x{ABCD})$ and $\xI{ABCD} \coloneqq \sI_4(\x{A:B:C:D})$.
}
\label{tab:HEI_N6}
\end{table}


Notice that the TF expression itself is quite compact; in fact there are over 300 orbits of $\nN=6$ HEIs whose TF expression has no more than 4 terms, and we found over 1400 orbits with 5 TF terms.
As an example, for $\Qn{23}$, the S-basis expression has 20 terms in S-basis, 24 terms in I-basis, but only 3 terms in the TF.   The corresponding \ino\ classification gets decomposed to $\Ink{10, 11, 3} = \Ink{4, 4, 1} + \Ink{4, 4, 1} + \Ink{2, 3, 1}$ in the latter form.  
In the diagrammatic representation introduced above, we have
\begin{center} 
    $\Qn{5}$ = \vTFtab[np]{5}{{AB,C,D},{A,C,E,D}} , \ \
    $\Qn{9}$ = \vTFtab[np]{5}{{AB,C,D},{B,D,E,C},{A,C,E,D}} , \ \
    $\Qn{10}$ = \vTFtab[np]{5}{{A,B,D},{B,C,E,A},{C,D,E,B},{A,C,E,D}} , \ \
    $\Qn{23}$ = \vTFtab[np]{6}{{A,BC,DE},{AD,BE,F},{A,B,CD,F}}  .
\end{center}

\paragraph{Purification symmetry:}
So far we only discussed composite subsystems made up from the $\nN$ elementary subsystems $\x{A}_i$, whose entanglement entropies are defined for any mixed state $\rho$.  
However, any mixed state $\rho$ on $\nN$ parties can be purified to a pure state on $\nN+1$ parties, where one of the parties is associated with the purifier (which we typically label by $\x{O}$); in the CFT we can identify the purifier with the complementary region (on a given Cauchy slice) to the others, and we can treat it on equal footing.  By purification symmetry, the entanglement entropy of any composite subsystem $\x{X}$ equals the entanglement entropy of the complementary subsystem $\overline{\x{X}}$, in other words $S(\x{X})=S(\overline{\x{X}})$, 
so we can always express any information quantity in the S-basis or I-basis, without explicitly referring to the purifier.  

The collection of all HEIs is of course invariant under the full permutation and purification symmetry $S_{\nN+1}$.  However, the S- and I- bases break the purification symmetry by explicitly choosing which party is the purifier (and hence not invoked in any of the terms).\footnote{\, 
    To remedy this, \cite{He:2019ttu} developed the K-basis, based on perfect tensor structures, whose key property is that it is fully $S_{\nN+1}$ invariant.  However, in this work we will not utilize this re-casting, partly because the full $S_{\nN+1}$ symmetry is anyway broken for the specific families of HEIs under consideration, but more importantly because the TF is more naturally obtained from the I-basis.
}  
Therefore we will be dealing with building blocks which are \emph{not} individually purification-invariant.  In particular, the $\sI_n$'s behave nontrivially under purification, and this feature is reflected in the fact that the \ino\ classification of a given $Q$ can change under purification (but of course not under any other permutation of the parties).  In particular, we can replace any argument of $I_3$ by the complement of all arguments,
\begin{equation}\label{eq:I3purif}
	\xtI{X:Y:Z} = I_3(\x{X:Y:}\overline{\x{XYZ}}) \ .
\end{equation}
For example, for $\nN=6$, $\xtI{A:B:C} = \xtI{A:B:DEFO}$, so the corresponding \ino\ classification would jump from $\Ink{1,0,0,0}$ to $\Ink{4, 6, 4, 1}$ when we replace $\x{O}$ by $\x{C}$ in the latter expression.  Nevertheless, for the unconditioned $I_3$ at least the sign remains the same under purification.  However this is no longer the case for the conditional $I_3$: here if we purify on the conditioned-on party, the sign actually flips:
\begin{equation}\label{eq:CI3purif}
	\xctI{X:Y:Z}{W} = -I_3(\x{X:Y:Z}|\overline{\x{XYZW}}) \ .
\end{equation}
This means that a general TF expression of the form \cref{eq:ctform}, when purified on any party contained in any of the $\x{W}_i$, would not remain TF-compatible unless all the wrong-sign terms get canceled.  The fact that according to \cref{conj:HEIsTF} the HEIs remain TF-compatible will then provide additional constraints on the TF; we return to this point in \cref{s:building} and \ref{s:discussion}.

\subsection{Two infinite families of HEIs}
\label{ss:inffam}

We now turn to the two infinite families of HEIs formulated by \cite{Czech:2023xed} (and shown to be primitive in \cite{Czech:2024rco}).  
The toric inequalities are labeled by a pair of odd integers $(m,n)$, and projective plane inequalities  are labeled by a single integer $m=n$.  In both families, the corresponding $\nN=m+n-1$ is odd (so at $\nN=6$ we only see the lifts).  To express these compactly, it will be convenient to readjust our notation (and specifically follow the notation of \cite{Czech:2023xed,Czech:2024rco}).  This is in fact a hybrid of the previously-introduced notation $\x{A,B,C,\ldots}$, with the purifier labeled by $\x{O}$ when we're treating the parties on equal footing, and the more enumerative notation $\x{A}_i$ with $i\in [\nN+1] \equiv \{1,2,\ldots,\nN,\nN+1\}$,  the last associated with the purifier.  In the case of the two infinite families in \cite{Czech:2024rco}, it is convenient to separate the regions into two sets, $\x{A}_i$ and $\x{B}_j$ where 
$i\in [m] $ and $j \in [n]$.\footnote{\,
	However, when presenting explicit examples of various members of these families, such as in \cref{tab:HEI_toric}, we will revert to the original $\x{A,B,C,\ldots}$ notation, because although it obscures the 2-set structure, the presentation is less cluttered and easier to read.   On the other hand, in order to manifest a different set of structural relations such as in \cref{eq:cg}, we employ different repartitionings.
}
Here by convention (unless noted otherwise), we treat $\x{B}_n$ as the purifier.  Furthermore, it will be convenient to introduce the integers $\mu$ and $\nu$, associated with the labels $m$ and $n$, defined as
\begin{equation}\label{eq:toriconvention}
    m=2\mu+1 \ \Longleftrightarrow \ \mu = \frac{m-1}{2} 
	\qquad \text{and} \qquad
	n=2\nu+1 \ \Longleftrightarrow \ \nu = \frac{n-1}{2} \ .
\end{equation}

These HEI families invoke consecutively indexed $k$-tuples of $\x{A}$  and $\x{B}$ regions, for which we define the shorthand
\begin{equation}\label{eq:Akdef}
	\xm{A}{i}{k} \coloneq \x{A}_i \x{A}_{i+1} \ldots \x{A}_{i+k-1}
	\qq{and}
	\xm{B}{j}{k} \coloneq \x{B}_j \x{B}_{j+1} \ldots \x{B}_{j+k-1}
\end{equation}
with the indices in each group identified cyclically: $\x{A}_{i+m} \simeq \x{A}_{i}$ (and similarly for $\x{B}$'s).
The toric family also invokes the further notational shortcut for the largest consecutively indexed minorities and smallest consecutively indexed majorities of $A$ and $B$ type regions:
$A_i^\pm \equiv \xm{A}{i}{(m\pm 1)/2}$ and similarly $B_j^\pm \equiv \xm{B}{j}{(n\pm 1)/2}$.  
With these notational simplifications in hand, we can now express the two HEI families from \cite{Czech:2023xed} quite compactly.

\paragraph{Toric HEI:}
The \qtor\ HEI family can be written as  (cf.\ \cite[eq.(4)]{Czech:2023xed}):
\begin{equation}\label{eq:Qtoric}
    \QT{m,n} \coloneq
    \sum_{i=1}^m \sum_{j=1}^n S_{A_i^{+} B_j^-}
    -
    \sum_{i=1}^m \sum_{j=1}^n S_{A_i^- B_j^-} \, - S_{A_1A_2\ldots A_m} 
    \ge 0
\end{equation}
where $m$ and $n$ are both odd.
It contains the \qcyc\ family when $n=1$;  in particular, MMI corresponds to $\QT{3,1}$.
Apart from the $(m,1)$ family, $\QT{3,3}$ is\footnote{\, 
	Here and below, when comparing two expressions for an information quantity, by ``is'' $Q$ we mean ``lies in the same orbit (under permutations and purifications) as'' $Q$, i.e., it is $Q$ up to relabeling the parties.
} $\Qn{10}$ from \cite{Hernandez-Cuenca:2023iqh}  (cf.\ \cref{tab:HEI_N6}), and $\QT{5,3}$ is the $\nN=7$ HEI from \cite[eq.(2.1)]{Czech:2022fzb}.
A representative selection of explicit examples for small $(m,n)$ members of this family, recast in our original notation and presented in the S-basis, I-basis, and the eventual TF, appears in \cref{tab:HEI_toric}.

\paragraph{Projective HEI:}
The \qproj\ HEI family can be written as (cf.\ \cite[eq.(6)]{Czech:2023xed}):
\begin{equation}\label{eq:Qproj}
    \QP{m} \coloneq
    \frac{1}{2} 
    \sum_{i,j=1}^{m} 
    \left(S_{A_i^{(j)} B_{i+j-1}^{(m-j)}} + S_{A_i^{(j)} B_{i+j}^{(m-j)}} \right)
    -
    \sum_{i,j=1}^{m} S_{A_i^{(j-1)} B_{i+j-1}^{(m-j)}}
    - S_{A_1 A_2 \ldots A_m} 
    \ge 0 
\end{equation}
or (as used in \cite[eq.(2.4)]{Czech:2024rco}, but for a different instance in the permutation orbit) 
\begin{equation}\label{eq:Qproj2}
    \QP{m} =
    \frac{1}{2} 
    \sum_{j=1}^{m-1} \sum_{i=1}^{m}
    \left(S_{A_i^{(j)} B_{i+j}^{(m-j)}} + S_{A_i^{(j)} B_{i+j+1}^{(m-j)}} \right)
    \,+\, (m-1)\,S_{A_1 A_2 \ldots A_m}
    -
    \sum_{i,j=1}^{m} S_{A_i^{(j-1)} B_{i+j}^{(m-j)}}
    \ge 0
\end{equation}
where $m>1$ can be odd or even.

One can easily check that $\QP{2}$ is MMI, and $\QP{3}$ is $\Qn{9}$ from \cite{Hernandez-Cuenca:2023iqh} (cf.\ \cref{tab:HEI_N6}). 
Explicit expressions of \cref{eq:Qproj} for $m=3,4,5$, rendered in  S-basis, I-basis, and the eventual TF, are presented in \cref{tab:HEI_proj}.

In the following two sections, we will rewrite these expressions in the TF, thereby explicitly demonstrating TF-compatibility.

\section{Toric HEI family}
\label{s:toric}

We start with the toric HEI family, \cref{eq:Qtoric}.  To provide further insight, rather than merely presenting the final TF, we detail how we actually arrived at the result, step by step.

We start by considering the $m=3$ family, the first few members of which are written out in  \cref{tab:qt3}.   

\begin{table}[htbp]
\begin{center}
\scriptsize
\begin{tabular}{|c|c|c|}
    \hline 
    \shadeR{label} & \shadeI{\ino\ classification} &  \shadeI{I-basis} \\
    \hline\hline
    $\QT{3,1}$ & \Ink{1,0,0,0} &  \hlorange{$-\xI{ABC}$}  \\
    $\QT{3,3}$ & \Ink{4,3,0,0}& \hlorange{$-\xI{ABC}$}$- \xI{ADE} - \xI{BDE} - \xI{CDE} $  \\
    & & $+\xI{ABDE} +\xI{ACDE} + \xI{BCDE}$ \\
    $\QT{3,5}$ & \Ink{10,15,6,0} &  \hlorange{$-\xI{ABC}$}$ -\xI{ADF} -\xI{AEF} -\xI{AEG} -\xI{BDF}  $  \\
     & & $  -\xI{BEF} -\xI{BEG} -\xI{CDF} -\xI{CEF} -\xI{CEG} $\\
     & & $+\xI{ABDF} +\xI{ABEF} +\xI{ABEG} + \xI{ACDF} + \xI{ACEF}  $ \\
     & & $ +\xI{ACEG} +\xI{ADEF}+\xI{AEFG}+\xI{BCDF}+\xI{BCEF} $ \\
     & & $+\xI{BCEG}+\xI{BDEF}+\xI{BEFG}+\xI{CDEF}+\xI{CEFG}$\\
     & & $ -\xI{ABDEF} -\xI{ABEFG} -\xI{ACDEF} -\xI{ACEFG} -\xI{BCDEF}-\xI{BCEFG} $\\
    $\QT{3,7}$ & \Ink{19, 42, 33, 9}& ... \\
    \hline 
\end{tabular}
\end{center}
\caption{ $\QT{3,n}$ series of \qtor\ inequalities, written out in the I-basis (right column), with the corresponding \ino\ classification (middle column) which counts the number of terms of different sizes, specifically $\{\#\sI_3,\#\sI_4 ,\#\sI_5,\#\sI_6 \}$.  The single term common to all members of this family is highlighted.}
\label{tab:qt3}
\end{table}
\noindent We observe two salient features, pertaining to the right and middle column respectively:
\begin{itemize}[nosep]
    \item The term $-\xI{ABC}$ always appears (as highlighted in \cref{tab:qt3})
    \item The \ino\ classification of the remaining terms can be decomposed into the \ino\ classification of conditional tripartite information terms with a rather suggestive pattern.
\end{itemize}
Let us examine the latter, utilizing the decomposition rule explained in  \cref{ss:ctform}.  In particular, with the schematic notation $I_{(1:1:k)}$ and $C_{(1:1:k|\ell)}$ introduced earlier to indicate the sizes of the arguments of the (conditional) tripartite informations, we list the \ino\ classification of the first few such terms in \cref{tab:inc}. 
\begin{table}[htbp]
\begin{center}
\scriptsize
\begin{tabular}{|c|c|}
    \hline 
    \shadeR{Information quantity} & \shadeI{\ino\ classification} \\
    \hline\hline
    $-I_{(1:1:1)}$ & \Ink{1,0,0}\\
    $-C_{(1:1:1|1)}$ & \Ink{1,1,0}\\
    $-C_{(1:1:1|2)}$ & \Ink{1,2,1}\\
    $-C_{(1,1,2|1)}$ & \Ink{2,3,1}\\
    \hline 
\end{tabular}
\end{center}
\caption{ The \ino\ classification of (conditional) tripartite information; notation explained in \cref{ss:ctform}.
}
\label{tab:inc}
\end{table}

Comparing these results with the \ino\ classification for the $\QT{3,n}$ family, we observe that the \ino\ classification in \cref{tab:qt3} can be decomposed as:
\begin{equation}\label{eq:indecomp}
\begin{aligned}
	\QT{3,1} & \sim -I_{(1:1:1)} \\
	\QT{3,3} & \sim -I_{(1:1:1)} - 3 C_{(1:1:1|1)} \\
	\QT{3,5} & \sim -I_{(1:1:1)} - 3 C_{(1:1:1|2)} - 3 C_{(1,1,2|1)} \\
	\QT{3,7} & \sim -I_{(1:1:1)} - 3 C_{(1:1:1|3)} - 3 C_{(1,1,2|2)} - 3 C_{(1,1,3|1)}
\end{aligned}
\end{equation}
The pattern is evident: in addition to the $\sI_3$ term, we sum 3 times the sum of $C_{(1:1:k|\ell)}$ terms with $k+\ell  = (n+1)/2 = \nu+1$ (where we used \cref{eq:toriconvention} for the last expression). This leads us to conjecture that all the $\QT{3,n}$ toric inequalities may be written in the form
\begin{equation}\label{eq:Q3nform}
	\QT{3,n} \sim  
	- I_{(1:1:1)} - 3 C_{(1:1:1|\nu) }- 3 C_{(1:1:2|\nu-1)} - ... - 3 C_{(1:1:\nu|1)} \ .
\end{equation}
By further matching the terms, we arrive at the following lemma:
\begin{lemma}\label{lmm:QT3n}
    The \qtor\ family with $m=3$ takes the form: 
    \begin{equation}\label{eq:Q3n}
            \QT{3,n} = -\ftI{\xs{A}{1}}{\xs{A}{2}}{\xs{A}{3}} - \sum_{i = 1}^{3}\sum_{k = 1}^{\nu} \fctI{\xs{A}{i}}{\xs{B}{\nu+k}}{\xm{B}{k}{\nu-k+1}}{\xs{A}{i+1}\xm{B}{\nu+1}{k-1}} \ .
    \end{equation}
\end{lemma}
In fact a closer observation reveals that \cref{lmm:QT3n} can be easily generalized, leading to a stronger lemma. Instead of merely relating 
$\QT{3,n}$ to $\QT{3,1} = - \ftI{\xs{A}{1}}{\xs{A}{2}}{\xs{A}{3}}$, it relates $\QT{m,n}$ to $\QT{m,1}$, a member of the \qcyc\ family:

\begin{lemma}\label{lmm:rela}
    The toric family labeled by $(m,n)$ can be re-expressed in the form:
    \begin{equation}\label{eq:Ttoc}
        \QT{m,n} = \QT{m,1} - \sum_{i = 1}^{m}\sum_{k = 1}^{\nu} \fctI{\xs{A}{i}}{\xs{B}{\nu+k}}{\xm{B}{k}{\nu-k+1}}{\xm{A}{i+1}{\mu}\xm{B}{\nu+1}{k-1}} \ .
    \end{equation}
\end{lemma}

We see that \cref{lmm:QT3n} is a special case of \cref{lmm:rela}, obtained by setting $m=3$. The proof of \cref{lmm:rela} follows from a direct calculation, detailed in \cref{ss:lem2pf}. This lemma shows that every \qtor\ inequality can be expressed as a combination of \qcyc\ inequalities and several conditional tripartite information terms with a specific pattern.

However, we are not done yet: while the double-sum term in \cref{eq:Ttoc} is already manifestly in the \ctform, the first term is not.  Nevertheless, the problem of constructing the TF of the \qtor\ inequality has now been reduced to finding the TF of the \qcyc\ inequalities, which correspond to the $\QT{m,1}$ family.  This is in fact the crux of the TF derivation. Curiously, the $\QT{1,n}$ family is much more straightforward: simply setting $m = 1$ in \cref{eq:Ttoc}, where $\QT{1,1} = 0$ and the region $\x{A}_1$ is the only $A$ region, immediately yields a simple TF expression:
\begin{equation}\label{eq:Q1n}
    \QT{1,n} = -\sum_{k = 1}^{\nu} \fctI{\xs{A}{1}}{\xs{B}{\nu+k}}{\xm{B}{k}{\nu-k+1}}{\xm{B}{\nu+1}{k-1}} \ .
\end{equation}
It is evident that the difference between the information quantities of $\QT{m,1}$ and $\QT{1,n}$ at $m=n$ lies in choosing region $\x{A}_1$ as the purifier instead of region $\x{B}_n$ and relabeling all the $B$ regions as $A$ regions.  But this by itself does not suffice, since as mentioned above, purifying an information quantity written in TF does not guarantee that it retains TF after the process. 
In particular, when the conditioned-on term contains the purifier, the purified expression flips sign (cf.~\cref{eq:CI3purif}), and we then need to rewrite the remaining terms in such a way as to cancel all the wrong-sign terms, leaving behind a true TF.  We postpone this exercise to  \cref{s:building}, where we can take advantage of systematically comparing with the \qproj\ family.   
The upshot is that the information quantity $\QT{n,1}$ can indeed be written in a \ctform.\footnote{\,
    There are in fact several alternate nice ways of writing this expression; while instructive, we don't need them in the main text, so we relegate their presentation to \cref{ss:qtorctform}.
}

\begin{lemma}\label{lmm:QTm1}
    The \qtor\ family labeled by $(m,1)$ takes the (tripartite) form:
    \begin{equation}\label{eq:Qm1}
    \begin{split}
        \QT{m,1} = & - \sum_{i = 1}^{\mu}\sum_{j = 1}^{i} \fctI{\xs{A}{\mu+j}}{\xs{A}{i}}{\xm{A}{\mu+i+1}{\mu-i+1}}{\xm{A}{i+1}{\mu-i}\xm{A}{\mu+j+1}{i-j}}\\
        & - \sum_{j = 2}^\mu \fctI{\xs{A}{\mu+j}}{\xm{A}{j}{\mu-j+1}}{\xm{A}{1}{j-1}}{\xm{A}{\mu+j+1}{\mu-j+1}} \ .
    \end{split}
    \end{equation}
\end{lemma}

The proof of \cref{lmm:QTm1} will be presented in \cref{ss:cycpur}.
Notice that while this form is manifestly more complicated than the $\QT{1,n}$ TF in \cref{eq:Q1n}, no region $B$ appears in the above expression.  This is because $n=1$ and so $\x{B}_1$ is the purifier.

Finally, by combining \cref{lmm:rela} and \cref{lmm:QTm1}, we see that the entire \qtor\ family, for every odd $m$ and $n$, can be recast in TF.  We state this result as our first key theorem, 
following our previous convention of $\x{B}_n$ being the purifier\footnote{\, 
    In \cref{thm:torct}, we explicitly present the \ctform\ with the region $\xs{B}{n}$ chosen as the purifier. Nevertheless, every element in the permutation-purification orbit of the \qtor\ inequality admits a \ctform\ of the same structure. Indeed, owing to the $D_m\times D_n$ dihedral symmetry, replacing the purifier by any region $\xs{B}{j}$ changes the displayed form only by a permutation of the region labels. Furthermore, the information quantities $\QT{m,n}$ and $\QT{n,m}$ have identical structures in the S-basis upon exchanging the $\x{A}$ and $\x{B}$ families. Consequently, the quantity $\QT{m,n}$ with any region $\xs{A}{i}$ chosen as the purifier admits a tripartite form of the same structure as that of $\QT{n,m}$ with a region $\xs{B}{j}$ chosen as the purifier.
}
and using the $(m,n) \leftrightarrow (\mu,\nu)$ conversion of \cref{eq:toriconvention}: 

\begin{thm}\label{thm:torct}
    The \qtor\ family labeled by $(m,n)$ is TF-compatible and takes the form:
    \begin{equation}\label{eq:torct}
    \begin{split}
        \QT{m,n} = & -\sum_{i = 1}^{m}\sum_{k = 1}^{\nu} \fctI{\xs{A}{i}}{\xs{B}{\nu+k}}{\xm{B}{k}{\nu-k+1}}{\xm{A}{i+1}{\mu}\xm{B}{\nu+1}{k-1}}\\
        & - \sum_{i = 1}^{\mu}\sum_{j = 1}^{i} \fctI{\xs{A}{\mu+j}}{\xs{A}{i}}{\xm{A}{\mu+i+1}{\mu-i+1}}{\xm{A}{i+1}{\mu-i}\xm{A}{\mu+j+1}{i-j}}\\
        & - \sum_{j = 2}^\mu \fctI{\xs{A}{\mu+j}}{\xm{A}{j}{\mu-j+1}}{\xm{A}{1}{j-1}}{\xm{A}{\mu+j+1}{\mu-j+1}} \ .
    \end{split}
    \end{equation}
\end{thm}
Note that when $m=1$ (so $\mu=0$), the last two lines of \cref{eq:torct} vanish,  when $m=3$ (so $\mu=1$), the last line vanishes, and when $n=1$ (so $\nu=0$), the first line vanishes.  When $m \ge 3$, the total number of (conditional) tripartite information terms is given by 
\begin{equation}\label{eq:toricNumTF}
\frac{m^2 + 4mn - 13}{8}\ ,
\end{equation}
which is remarkably compact:  The S-basis representation has $(2mn+1)$ terms, while the I-basis representation scales exponentially with $m$ and $n$; the numerical values are presented in \cref{tab:ctterm} for each type of expression for first few values of $m$ and $n$.
\begin{table}[htbp]
\begin{center}
\scriptsize
\begin{tabular}{|c||c|c|c||c|c|c||c|c|c||c|c|c||c|c|c|}
    \hline
    \shadeG{} 
    &\multicolumn{3}{c||}{ \shadeG{$n = 1$}} 
    &\multicolumn{3}{c||}{ \shadeG{$n = 3$}} 
    &\multicolumn{3}{c||}{ \shadeG{$n = 5$}} 
    &\multicolumn{3}{c||}{ \shadeG{$n = 7$}} 
    &\multicolumn{3}{c|}{ \shadeG{$n = 9$}} 
      \\ \hline \hline
    \shadeG{$m = 1$} 
    &  \multicolumn{3}{c||}{} & 7 & \shadeI{1} & \shadeTF{1} & 11 & \shadeI{5} & \shadeTF{2} & 15 & \shadeI{17} & \shadeTF{3}
    & 19 & \shadeI{49} & \shadeTF{4}
     \\ \hline
    \shadeG{$m = 3$} 
    & 7 & \shadeI{1} & \shadeTF{1} & 19 & \shadeI{7} & \shadeTF{4} & 31 & \shadeI{31} & \shadeTF{7} & 43 & \shadeI{103} & \shadeTF{10}
    & 55 & \shadeI{295} & \shadeTF{13}
     \\ \hline
    \shadeG{$m = 5$} 
    & 11 & \shadeI{11} & \shadeTF{4} & 31 & \shadeI{31} & \shadeTF{9} & 51 & \shadeI{111} & \shadeTF{14} & 71 & \shadeI{351} & \shadeTF{19}
    & 91 & \shadeI{991} & \shadeTF{24}
     \\ \hline
    \shadeG{$m = 7$} 
    & 15 & \shadeI{71} & \shadeTF{8} & 43 & \shadeI{127} & \shadeTF{15} & 71 & \shadeI{351} & \shadeTF{22} & 99 & \shadeI{1023} & \shadeTF{29}
    & 127 & \shadeI{2815} & \shadeTF{36}
     \\ \hline
     \shadeG{$m = 9$} 
    & 19 & \shadeI{367} & \shadeTF{13} & 55 & \shadeI{511} & \shadeTF{22} & 91 & \shadeI{1087} & \shadeTF{31} & 127 & \shadeI{2815} & \shadeTF{40}
    & 163 & \shadeI{7423} & \shadeTF{49}
     \\ 
    \hline
\end{tabular}
\end{center}
\caption{The number of terms in $\QT{m,n}$ when expressed in the \ctform\ \cref{eq:torct} (highlighted in pink), as compared to the total number of terms in the S-basis (shown in white) and the I-basis (shown in blue). 
}
\label{tab:ctterm}
\end{table}
Interestingly, 
we see there is a strong asymmetry between the $A$ terms and the $B$ terms (correspondingly between $m$ and $n$), which is not apparent in its S-basis representation \cref{eq:Qtoric}.  In particular, adding more $B$ terms only augments the first line of \cref{eq:torct} whereas adding more $A$ terms augments all three lines, which is reflected in \cref{eq:toricNumTF}.
We return to a more refined counting in \cref{s:discussion}, cf.~\cref{tab:incT}.

\paragraph{Explicit expressions:}
Now that we have presented the \qtor\ family in TF compactly, let us pause to examine the actual form in more detail.  This offers an independent insight into the  structure of $\QT{m,n}$, and as we will see in the next section, reveals interesting connections to the \qproj\ family.
The \cref{tab:HEI_toric} lists the first few instances of $\QT{m,n}$ for small $(m,n)$.  In the interest of compactness of presentation the expressions are rewritten in the $\x{A,B,C},\ldots$ notation previously employed in \cref{tab:HEI_N6}.

%
\begin{table}[htbp] 
\begin{center}
\scriptsize
\begin{tabular}{| c | c | }
\hline
label &  $\QT{m,n}$ information quantity \\ 
\hline
\hline
$\QT{1,5}$
    &  $ -\xS{A}-\xS{BC}-\xS{CD}-\xS{DE}-\xS{ABCD}-\xS{ACDE}+\xS{ABC}+\xS{ACD}+\xS{ADE}+\xS{BCD}+\xS{CDE} $ 
\\ 
    & \shadeI{
    $ +\xI{ABCD}+\xI{ACDE}-\xI{ABD}-\xI{ACD}-\xI{ACE} $ 
    } 
\\ 
    & \shadeTF{
    $-\tI{A:D:BC}-\ctI{A:E:C}{D}$ 
    } 
\\ \hline
$\QT{5,1}$
    &   $ -\xS{AB}-\xS{AE}-\xS{BC}-\xS{CD}-\xS{DE}-\xS{ABCDE}+\xS{ABC}+\xS{ABE}+\xS{ADE}+\xS{BCD}+\xS{CDE} $ 
\\ 
    & \shadeI{
    $ -\xI{ABCDE}+\xI{ABCD}+\xI{ABCE}+\xI{ABDE}+\xI{ACDE}+\xI{BCDE}-\xI{ABD}-\xI{ACD}-\xI{ACE}-\xI{BCE}-\xI{BDE} $ 
    } 
\\ 
    & \shadeTF{
    $-\ctI{C:A:DE}{B}-\ctI{C:B:E}{D}-\tI{D:B:E}-\ctI{D:B:A}{E}$ 
    } 
\\ \hline
$\QT{3,3}$
    &  \makecell{
        $ -\xS{AD}-\xS{AE}-\xS{BD}-\xS{BE}-\xS{CD}-\xS{CE}-\xS{ABC}-\xS{ABDE}-\xS{ACDE}-\xS{BCDE} $ \\
        $ +\xS{ABD}+\xS{ABE}+\xS{ACD}+\xS{ACE}+\xS{ADE}+\xS{BCD}+\xS{BCE}+\xS{BDE}+\xS{CDE} $ }
\\ 
    & \shadeI{
    $ +\xI{ABDE}+\xI{ACDE}+\xI{BCDE}-\xI{ABC}-\xI{ADE}-\xI{BDE}-\xI{CDE} $ 
    } 
\\ 
    & \shadeTF{
    $-\ctI{A:E:D}{B}-\ctI{B:E:D}{C}-\ctI{C:E:D}{A}-\tI{B:A:C}$ 
    } 
\\ \hline
$\QT{3,5}$
    &  \makecell{
        $ -\xS{ABC}-\xS{ADE}-\xS{AEF}-\xS{AFG}-\xS{BDE}-\xS{BEF}-\xS{BFG}-\xS{CDE}-\xS{CEF}-\xS{CFG}-\xS{ABDEF} $ \\
        $ -\xS{ABEFG}-\xS{ACDEF}-\xS{ACEFG}-\xS{BCDEF}-\xS{BCEFG}+\xS{ABDE}+\xS{ABEF}+\xS{ABFG}+\xS{ACDE}+\xS{ACEF} $ \\
        $ +\xS{ACFG}+\xS{ADEF}+\xS{AEFG}+\xS{BCDE}+\xS{BCEF}+\xS{BCFG}+\xS{BDEF}+\xS{BEFG}+\xS{CDEF}+\xS{CEFG} $}
\\ 
    & \shadeI{\makecell{
    $-\xI{ABDEF}-\xI{ABEFG}-\xI{ACDEF}-\xI{ACEFG}-\xI{BCDEF}-\xI{BCEFG}+\xI{ABDF}+\xI{ABEF}+\xI{ABEG}+\xI{ACDF}+\xI{ACEF}$ \\
    $+\xI{ACEG}+\xI{ADEF}+\xI{AEFG}+\xI{BCDF}+\xI{BCEF}+\xI{BCEG}+\xI{BDEF}+\xI{BEFG}+\xI{CDEF}+\xI{CEFG}$ \\
    $-\xI{ABC}-\xI{ADF}-\xI{AEF}-\xI{AEG}-\xI{BDF}-\xI{BEF}-\xI{BEG}-\xI{CDF}-\xI{CEF}-\xI{CEG}$ 
    } }
\\ 
  & \shadeTF{ 
  $-\ctI{A:F:DE}{B}-\ctI{A:G:E}{BF}-\ctI{B:F:DE}{C}-\ctI{B:G:E}{CF}-\ctI{C:F:DE}{A}-\ctI{C:G:E}{AF}-\tI{B:A:C}$}
\\  \hline
$\QT{3,7}$
    &  \makecell{
        $ -\xS{ABC}-\xS{ADEF}-\xS{AEFG}-\xS{AFGH}-\xS{AGHI}-\xS{BDEF}-\xS{BEFG}-\xS{BFGH}-\xS{BGHI}-\xS{CDEF} $ \\
        $ -\xS{CEFG}-\xS{CFGH}-\xS{CGHI}-\xS{ABDEFG}-\xS{ABEFGH}-\xS{ABFGHI}-\xS{ACDEFG}-\xS{ACEFGH}-\xS{ACFGHI} $ \\
        $ -\xS{BCDEFG}-\xS{BCEFGH}-\xS{BCFGHI}+\xS{ABDEF}+\xS{ABEFG}+\xS{ABFGH}+\xS{ABGHI}+\xS{ACDEF}+\xS{ACEFG} $ \\
        $ +\xS{ACFGH}+\xS{ACGHI}+\xS{ADEFG}+\xS{AEFGH}+\xS{AFGHI}+\xS{BCDEF}+\xS{BCEFG}+\xS{BCFGH}+\xS{BCGHI} $ \\
        $ +\xS{BDEFG}+\xS{BEFGH}+\xS{BFGHI}+\xS{CDEFG}+\xS{CEFGH}+\xS{CFGHI} $ }
\\ 
    & \shadeI{\makecell{
    $ \xI{ABDEFG}+\xI{ABEFGH}+\xI{ABFGHI}+\xI{ACDEFG}+\xI{ACEFGH}+\xI{ACFGHI}+\xI{BCDEFG}+\xI{BCEFGH} +\xI{BCFGHI}$ \\
    $ -\xI{ABDEG}-\xI{ABDFG}-\xI{ABEFG}-\xI{ABEFH}-\xI{ABEGH}-\xI{ABFGH}-\xI{ABFGI}-\xI{ABFHI} -\xI{ACDEG}-\xI{ACDFG}$ \\
    $ -\xI{ACEFG}-\xI{ACEFH}-\xI{ACEGH}-\xI{ACFGH}-\xI{ACFGI}-\xI{ACFHI}-\xI{ADEFG} -\xI{AEFGH}-\xI{AFGHI}-\xI{BCDEG}$ \\
    $ -\xI{BCDFG}-\xI{BCEFG}-\xI{BCEFH}-\xI{BCEGH}-\xI{BCFGH}-\xI{BCFGI} -\xI{BCFHI}-\xI{BDEFG}-\xI{BEFGH}-\xI{BFGHI}$ \\
    $ -\xI{CDEFG}-\xI{CEFGH}-\xI{CFGHI}+\xI{ABDG}+\xI{ABEG} +\xI{ABEH}+\xI{ABFG}+\xI{ABFH}+\xI{ABFI}+\xI{ACDG}+\xI{ACEG}$ \\
    $ +\xI{ACEH}+\xI{ACFG}+\xI{ACFH}+\xI{ACFI} +\xI{ADEG}+\xI{ADFG}+\xI{AEFG}+\xI{AEFH}+\xI{AEGH}+\xI{AFGH}+\xI{AFGI}+\xI{AFHI}$ \\
    $ +\xI{BCDG}+\xI{BCEG} +\xI{BCEH}+\xI{BCFG}+\xI{BCFH}+\xI{BCFI}+\xI{BDEG}+\xI{BDFG}+\xI{BEFG}+\xI{BEFH}+\xI{BEGH}$ \\
    $ +\xI{BFGH} +\xI{BFGI}+\xI{BFHI}+\xI{CDEG}+\xI{CDFG}+\xI{CEFG}+\xI{CEFH}+\xI{CEGH}+\xI{CFGH}+\xI{CFGI}+\xI{CFHI} $ \\
    $ -\xI{ABC}-\xI{ADG}-\xI{AEG}-\xI{AEH}-\xI{AFG}-\xI{AFH}-\xI{AFI}-\xI{BDG}-\xI{BEG}-\xI{BEH}-\xI{BFG}  -\xI{BFH}-\xI{BFI}$ \\
    $-\xI{CDG}-\xI{CEG}-\xI{CEH}-\xI{CFG}-\xI{CFH}-\xI{CFI} $ 
    } }
\\ 
    & \shadeTF{\makecell{
    $-\ctI{A:G:DEF}{B}-\ctI{A:H:EF}{BG}-\ctI{A:I:F}{BGH}-\ctI{B:G:DEF}{C}-\ctI{B:H:EF}{CG}-\ctI{B:I:F}{CGH}$ \\
    $-\ctI{C:G:DEF}{A}-\ctI{C:H:EF}{AG}-\ctI{C:I:F}{AGH}-\tI{B:A:C}$
    } }
\\ \hline
 \hline 
\end{tabular}
\end{center}
\caption{
Explicit expressions for $\QT{m,n}$ for small $(m,n)$ in the S-basis (white background), I-basis (blue background), and the \ctform\ (pink background). 
The notational shorthand and conventions are the same as in \cref{tab:HEI_N6} and \cite{Hernandez-Cuenca:2023iqh}.
We use the  (conventional) notation for the subsystems $\x{A},\x{B}\ldots$ for sake of compactness, though it obscures  the relational simplification of \eqref{eq:torct}, and in particular becomes harder to compare across different $(m,n)$.
}
\label{tab:HEI_toric}
\end{table}
We see that the \ctform\ is quite involved; on the other hand, it follows a clear pattern.  To get a better sense of the general pattern, it is useful to consider the diagrammatic representation.  
Instead of presenting the same set of $\QT{m,n}$ as in \cref{tab:HEI_toric} and in order to examine larger expressions, we present a more systematic family in \cref{tab:QTgraphicalrep}:  we fix $\nN=9$ and display all the $(m,n)$ possibilities with $m+n-1=9$.  
\\
\begin{table}[htbp]
    \begin{center}
    \vTFtab[np, cell size=0.26,box size=0.13, header font=\sffamily\fontsize{5}{12}\selectfont,
    title=$\QT{1,9}$]{9}{{A, F, BCDE}, {A, G, CDE, F}, {A, H, DE, FG}, {A, I, E, FGH}}
    \qquad
    \vTFtab[np, cell size=0.26,box size=0.13, header font=\sffamily\fontsize{5}{12}\selectfont,
    title=$\QT{3,7}$]{9}{{A, G, DEF, B}, {A, H, EF, BG}, {A, I, F, BGH}, {B, G, DEF, C}, {B, H, EF, CG}, {B, I, F, CGH}, {C, G, DEF, A}, {C, H, EF, AG}, {C, I, F, AGH}, {B, A, C}}
    \qquad
    \vTFtab[np, cell size=0.26,box size=0.13, header font=\sffamily\fontsize{5}{12}\selectfont,
    title=$\QT{5,5}$]{9}{{A, H, FG, BC}, {A, I, G, BCH}, {B, H, FG, CD}, {B, I, G, CDH}, {C, H, FG, DE}, {C, I, G, DEH}, {D, H, FG, AE}, {D, I, G, AEH}, {E, H, FG, AB}, {E, I, G, ABH}, {C, A, DE, B}, {C, B, E, D}, {D, B, E}, {D, B, A, E}}
    \qquad
    \vTFtab[np, cell size=0.26,box size=0.13, header font=\sffamily\fontsize{5}{12}\selectfont,
    title=$\QT{7,3}$]{9}{{A, I, H, BCD}, {B, I, H, CDE}, {C, I, H, DEF}, {D, I, H, EFG}, {E, I, H, AFG}, {F, I, H, ABG}, {G, I, H, ABC}, {D, A, EFG, BC}, {D, B, FG, CE}, {E, B, FG, C}, {D, C, G, EF}, {E, C, G, F}, {F, C, G}, {E, BC, A, FG}, {F, C, AB, G}}
    \qquad
    \vTFtab[np, cell size=0.26,box size=0.13, header font=\sffamily\fontsize{5}{12}\selectfont,
    title=$\QT{9,1}$]{9}{{E, A, FGHI, BCD}, {E, B, GHI, CDF}, {F, B, GHI, CD}, {E, C, HI, DFG}, {F, C, HI, DG}, {G, C, HI, D}, {E, D, I, FGH}, {F, D, I, GH}, {G, D, I, H}, {H, D, I}, {F, BCD, A, GHI}, {G, CD, AB, HI}, {H, D, ABC, I}}
    \end{center}
\caption{Diagrammatic representation of instances of $\QT{m,n}$ for $\nN=m+n-1=9$.}
\label{tab:QTgraphicalrep}
\end{table}

\paragraph{Singleton-tripartite form:}
Recall that the \ctform\ of a given TF-compatible information quantity is, in general, not unique.  As explained in \cref{ss:ctform}, much (though not all) of this vast redundancy is removed by considering the singleton-tripartite form (STF), where in each (conditional) tripartite information term, the three principal (unconditioned) arguments are singletons.
The conversions are obtained by generalizing the iterated structure of \cref{eq:multTFs}, as exemplified by \cref{eq:toSFT} and \cref{eq:toSFTcond}.  A single cardinality-$k$ argument (for convenience expressed as $\xs{A}{1}\xs{A}{2}\ldots \xs{A}{k} = \xm{A}{1}{k}$, with all other arguments denoted by $\xs{B}{j}$ and coming along for the ride) expands into $k$ STF terms:
\begin{equation}\label{eq:TFtoSTFgen}
    \ftI{\xs{B}{1}}{\xs{B}{2}}{\xm{A}{1}{k}}
    = \sum_{i=1}^{k} \fctI{\xs{B}{1}}{\xs{B}{2}}{\xs{A}{i}}{\xm{A}{i+1}{k-i}} \ .
\end{equation}

One of the main advantages of STF is that each $C_{(1:1:1|\ell)}$ has an independent \ino\ classification, which allows us to extract a unique \cno\ classification, characterizing how many $C_{(1:1:1|\ell)}$ terms there are at given $\ell$.  Since the building block classification follows the rows of the Pascal's triangle, as illustrated in \cref{tab:sing} (extended from \cref{tab:inc}), the conversion is quite simple (cf.\ \cref{eq:i2c}).

\begin{table}[htbp]  
\begin{center}
\scriptsize
\begin{tabular}{|c|c|c|}
    \hline 
    \shadeR{singleton-tripartite form} & \shadeI{\ino\ classification} &  \shadeR{\cno\ classification} \\
    \hline\hline
    $-I_{(1:1:1)}$ & \Ink{1,0,0,0,0} &
    \Cnk{1,0,0,0,0}
    \\
    $-C_{(1:1:1|1)}$ & \Ink{1,1,0,0,0} &
    \Cnk{0,1,0,0,0}
    \\
    $-C_{(1:1:1|2)}$ & \Ink{1,2,1,0,0} &
    \Cnk{0,0,1,0,0}
    \\
    $-C_{(1:1:1|3)}$ & \Ink{1,3,3,1,0} &
    \Cnk{0,0,0,1,0}
    \\
    $-C_{(1:1:1|4)}$ & \Ink{1,4,6,4,1} &
    \Cnk{0,0,0,0,1}
    \\
    \hline 
\end{tabular}
\end{center}
\caption{The \ino\ and \cno\ classification of the STF building blocks.
}
\label{tab:sing}
\end{table}

For completeness, we now present the information quantities of $ \QT{1,n}$, $ \QT{3,n}$, $ \QT{m,1}$, and  $ \QT{m,n}$ (listed in order of increasing complexity),  all in STF.  
For each, we specify the \cno\ classification and the total number of terms.  
\begin{itemize} 
\item 
    The $ \QT{1,n} $ family can be expanded as  (cf.~\cref{eq:Q1n}) 
    \begin{equation}
          \QT{1,n} = - \sum_{k=1}^{\nu} \sum_{j=\nu - k + 1}^{\nu}
        \fctI{\xs{A}{1}}{\xs{B}{j}}{\xs{B}{j+k}}{\xm{B}{j+1}{k-1}} \ .
    \end{equation}
    The summands have the form $C_{(1:1:1|k-1)}$, and the inner sum gives $k$ such terms, so that $c_{k-1}=k$.  The outer sum then collects these into the \cno\ 
    $\Cnk{1,2,\ldots,\nu}$ with remaining entries (starting with $c_\nu$) all vanishing. Summing these up, the expression has $  (\nu +1)\nu/2 $ terms in total.

\item 
    The $ \QT{3,n} $ family is then expanded into (cf.~\cref{eq:Q3n}) 
    \begin{equation}
        \QT{3,n} =   - \ftI{\xs{A}{1}}{\xs{A}{2}}{\xs{A}{3}} 
        -\sum_{i=1}^{3} \sum_{k=1}^{\nu} \sum_{j=\nu - k + 1}^{\nu}
       \fctI{\xs{A}{i}}{\xs{B}{j}}{\xs{B}{j+k}}{\xs{A}{i+1} \, \xm{B}{j+1}{k-1}} \ ,
    \end{equation}
    with \cno\ $\Cnk{1,3,6,9,\ldots,3\nu}$ with
    $ 1+ 3(\nu +1)\nu/2 $ terms in total.

\item 
    The $ \QT{m,1} $ family is (cf.~\cref{eq:Qm1})\footnote{\, 
    The singleton tripartite form follows directly from the expression \cref{eq:Qm12a} in \cref{ss:qtorctform}.
    }
    \begin{equation}
    \label{eq:CycSTF}
    \begin{aligned}
    \QT{m,1}
    ={}&
    -\sum_{k=1}^{\mu}
     \sum_{i=1}^{\mu-k+1}
     \sum_{j=k}^{i+k-1}
    \fctI
    {\xs{A}{j}}
    {\xs{A}{\mu+k}}
    {\xs{A}{\mu+k+i}}
    {
     \xm{A}{\mu+1}{k-1}
     \xm{A}{\mu+k+i+1}{\mu-i}
     \xm{A}{j+1}{k+i-j-1}
    }
    \\
    &-
    \sum_{i=1}^{\mu-1}
    \sum_{j=i+1}^{\mu}
    \sum_{k=1}^{i}
    \fctI
    {\xs{A}{k}}
    {\xs{A}{\mu+i+1}}
    {\xs{A}{j}}
    {
     \xm{A}{\mu+1}{i}
     \xm{A}{j+1}{\mu-j}
     \xm{A}{k+1}{i-k}
    }.
    \end{aligned}
    \end{equation}
    which has $(2\mu+1)(\mu +1)\mu/6 = (m^2 - 1)m/24 $ terms in total.  A closer inspection of the \cno\ classification reveals that it behaves as \cno\ $\Cnk{1,3,6,..., \frac{\mu(\mu+1)}{2},...,6,3,1}$ or in a more general form $\Cnk{{2 \choose 2},{3 \choose 2}, ..., {\mu+1 \choose 2},..., {3 \choose 2},{2 \choose 2} }$.

\item 
    Finally, the full $ \QT{m,n} $ family is (cf.\cref{eq:torct}):
    \begin{equation}
    \begin{aligned}
    \label{eq:QtorSTF}
    \QT{m,n}
    ={}&
    -\sum_{i=1}^{m} 
     \sum_{k=1}^{\nu} 
     \sum_{j=\nu - k + 1}^{\nu}
    \fctI
    {\xs{A}{i}}
    {\xs{B}{j}}
    {\xs{B}{j+k}}
    {\xm{A}{i+1}{\mu} \, \xm{B}{j+1}{k-1}}\\
    &
    -\sum_{k=1}^{\mu}
     \sum_{i=1}^{\mu-k+1}
     \sum_{j=k}^{i+k-1}
    \fctI
    {\xs{A}{j}}
    {\xs{A}{\mu+k}}
    {\xs{A}{\mu+k+i}}
    {
     \xm{A}{\mu+1}{k-1}
     \xm{A}{\mu+k+i+1}{\mu-i}
     \xm{A}{j+1}{k+i-j-1}
    }
    \\
    &-
    \sum_{i=1}^{\mu-1}
    \sum_{j=i+1}^{\mu}
    \sum_{k=1}^{i}
    \fctI
    {\xs{A}{k}}
    {\xs{A}{\mu+i+1}}
    {\xs{A}{j}}
    {
     \xm{A}{\mu+1}{i}
     \xm{A}{j+1}{\mu-j}
     \xm{A}{k+1}{i-k}
    }.
    \end{aligned}
    \end{equation}
    where the last two lines replicate $\QT{m,1}$ from \cref{eq:CycSTF}.  
    The first line has $c_\ell = m(\ell-\mu+1)$ for $\mu \leq \ell \leq \mu+\nu -1$, and 0 otherwise. 
    This gets combined with \cno\ of $\QT{m,1}$ which takes the values ${\ell+2 \choose 2}$ for $0\leq \ell\leq \mu-1$ and ${2\mu-\ell \choose 2}$ for $ \mu\leq \ell\leq 2\mu-2$ and 0 otherwise. Therefore the full STF for $\QT{m,n}$ in \cref{eq:QtorSTF} has $ \frac{(2\mu + 1) \, (\nu + 1)\, \nu }{2} + \frac{(2\mu + 1) \, (\mu + 1)\, \mu}{6} $ terms.
\end{itemize} 

\paragraph{Iterated refinement interpretation of the \qcyc\ HEIs:}
We close this section by exhibiting a completely different but tantalizingly simple form for the \qcyc\ family of HEIs.
Recall that the TF conversion of this $ \QT{m,1} $ subfamily, out of the entire $ \QT{m,n} $ toric family, was the most difficult part -- which may seem rather surprising, given its astonishing simplicity in the S-basis.  The disproportionately large size of the expression in the I-basis (cf.\ the blue column for $n=1$ in \cref{tab:ctterm}) may be part of the explanation,  but this does not mean that grouping terms into $I_3$'s is unrevealing.  In fact, one \emph{can} write this family in a remarkably simple form using just tripartite information terms.\footnote{\, 
    The reason this simple form is not a valid TF is that when converted into $Q \ge 0$ form, each $Q$ contains some wrong-sign ($+I_3$ instead of $-I_3$) terms.
}   In particular, 
we can express the whole family as a sequence of inequalities in a rather suggestive form:\footnote{\, 
    We use the $\text{C}_{i}$, $\text{D}_{i}$ notation to avoid confusion with the $\text{A}_{i}$, $\text{B}_{i}$ regions; to obtain it, starting from the $\QT{1,n}$ family, we replace $\text{A}_{1}$ by the purifier $\text{O}$, and we split the $(n-1)$ $\text{B}_{j}$ terms in half, regrouping them into the $\text{C}_i$ and $\text{D}_i$ terms.  
    Note that the interpretation of $\text{O}$ in each line depends on $\nN$; each pair of lines corresponds to different $\nN$, with the whole sequence generating the whole $ \QT{m,1} $ family.
}
\begin{equation}\label{eq:cg}
\begin{split}
    0 \leq &- \ftI{\x{O}}{\xs{C}{1}}{\xs{D}{1}}\\
    \leq & - \ftI{\x{O}}{\xs{C}{1}\xs{C}{2}}{\xs{D}{1}} - \ftI{\x{O}}{\xs{C}{1}}{\xs{D}{1}\xs{D}{2}} \\
    \leq & - \ftI{\x{O}}{\xs{C}{1}\xs{C}{2}\xs{C}{3}}{\xs{D}{1}} - \ftI{\x{O}}{\xs{C}{1}\xs{C}{2}}{\xs{D}{1}\xs{D}{2}}- \ftI{\x{O}}{\xs{C}{1}}{\xs{D}{1}\xs{D}{2}\xs{D}{3}} \\
    \leq & \quad \cdots
\end{split}
\end{equation}
To make contact with the usual form, for example for the third inequality, we would replace $\x{C}_3 \to \x{A}$, $\x{C}_2 \to \x{B}$, $\x{C}_1\to \x{C}$, $\x{D}_1 \to \x{D}$, $\x{D}_2 \to \x{E}$, $\x{D}_3 \to \x{F}$, expand out the $I_3$'s, and replace every term containing $\x{O}$ by its complement, to get
$ \xS{ABCD} +  \xS{BCDE} +  \xS{CDEF} +  \xS{DEFG} +  \xS{EFGA} +  \xS{FGAB} +  \xS{GABC} 
\ge \xS{ABC} + \xS{BCD} +  \xS{CDE} +  \xS{DEF} +  \xS{EFG} +  \xS{FGA} +  \xS{GAB} +  \xS{ABCDEFG} $.

\begin{figure}
    \centering
    \includegraphics[width=0.9\linewidth]{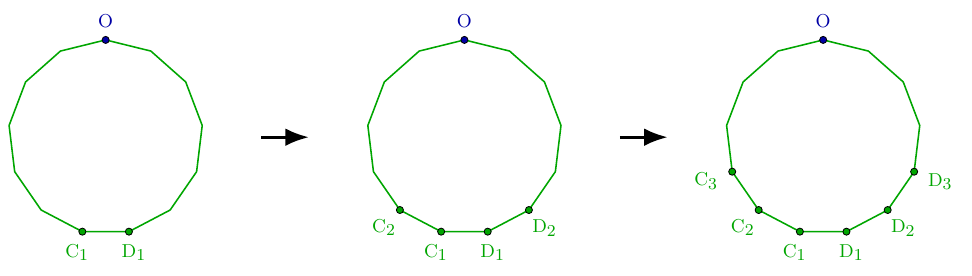}
    \caption{A schematic illustration of the iterated refinement interpretation of the \qcyc\ family.}
    \label{fig:cg}
\end{figure}

The form \cref{eq:cg} seems rather suggestive.  Notice that the $n$th line invokes $n$ $\x{C}$-regions and $n$ $\x{D}$-regions, occurring in a fixed ordering indicated in \cref{fig:cg}.  There are $n$ ways of partitioning these $2n$ consecutive $\x{CD}$ regions into ($\x{C}:\x{D}$) arguments with fixed composite size $n$ and preserved order, and we sum over all of these.  Each successive refinement is then bounded by the previous one.

This process seems somewhat reminiscent of successive fine-graining, and one might be tempted to attribute the monotonicity to this.  However, we are not averaging, but rather summing over the possibilities. On the other hand, because the total coefficient increases (i.e.\ the $n$th line has $n$ terms) and each $-I_3\ge0$, the inequalities might not seem so surprising.  What is perhaps more surprising, given this observation, is that each inequality still remains ``primitive'' (i.e.\ non-redundant, cf.~\cref{fn:primitive}).  Consequently, averaging rather than summing over the possibilities would in fact not have given a valid inequality.

\section{Projective HEI family}
\label{s:proj}

We now turn to the \qproj\ family of inequalities, and again indicate how we went about constructing the corresponding \ctform.  Recall from \cref{ss:inffam} that  two distinct S-basis forms have been previously presented in the literature \cite{Czech:2023xed,Czech:2024rco}, \cref{eq:Qproj} and \cref{eq:Qproj2}, which are related by permutations and purifications (in particular the $\x{B}$ terms get cyclically shifted by 1).  Here it will be convenient to use the first form, i.e.\ define $\QP{m}$ as in \cref{eq:Qproj},
\[
    \QP{m} \coloneq
    \frac{1}{2} 
    \sum_{i,j=1}^{m} 
    \left(S_{A_i^{(j)} B_{i+j-1}^{(m-j)}} + S_{A_i^{(j)} B_{i+j}^{(m-j)}} \right)
    -
    \sum_{i,j=1}^{m} S_{A_i^{(j-1)} B_{i+j-1}^{(m-j)}}
    - S_{A_1 A_2 \ldots A_m} \ .
\]
Interestingly, even though the S-basis expression for the \qproj\ family appears more complicated than that of the \qtor\ one (cf.~\cref{eq:Qtoric}), it will turn out that the TF conversion is actually simpler for the \qproj\ family.

We proceed similarly as in \cref{s:toric}.  We start by examining $\QP{m}$ for small $m$, rewritten in the I-basis.  
In \cref{tab:qp} we list the first few terms along with their \ino\ classification.
\begin{table}[htbp]
\begin{center}
\scriptsize
\begin{tabular}{|c|c|c|}
    \hline 
    \shadeR{label} & \shadeI{\ino\ classification} &  \shadeI{I-basis} \\
    \hline\hline
    $\QP{2}$ & \Ink{1,0,0,0} &  \hlorange{$-\xI{\xx{A}{1}\xx{A}{2}\xx{B}{1}}$}  \\
    $\QP{3}$ & \Ink{4,3,0,0}& \hlorange{$-\xI{\xx{A}{1}\xx{A}{2}\xx{B}{1}}$} \\
     & & \hlcyan{$-\xI{\xx{A}{1}\xx{A}{3}\xx{B}{1}}-\xI{\xx{A}{1}\xx{A}{3}\xx{B}{2}}-\xI{\xx{A}{2}\xx{A}{3}\xx{B}{2}} $}  \\
     & & \hllime{$+\xI{\xx{A}{1}\xx{A}{2}\xx{A}{3}\xx{B}{1}} +\xI{\xx{A}{1}\xx{A}{2}\xx{A}{3}\xx{B}{2}} +\xI{\xx{A}{1}\xx{A}{3}\xx{B}{1}\xx{B}{2}}$} \\
    $\QP{4}$ & \Ink{10,15,6,0} &  \hlorange{$-\xI{\xx{A}{1}\xx{A}{2}\xx{B}{1}}$}\\
     & & \hlcyan{$-\xI{\xx{A}{1}\xx{A}{3}\xx{B}{1}}-\xI{\xx{A}{1}\xx{A}{3}\xx{B}{2}}-\xI{\xx{A}{2}\xx{A}{3}\xx{B}{2}} $}  \\
     & & $ -\xI{\xx{A}{1}\xx{A}{4}\xx{B}{1}}-\xI{\xx{A}{1}\xx{A}{4}\xx{B}{2}} -\xI{\xx{A}{1}\xx{A}{4}\xx{B}{3}}$\\
     & & $ -\xI{\xx{A}{2}\xx{A}{4}\xx{B}{2}}-\xI{\xx{A}{2}\xx{A}{4}\xx{B}{3}}-\xI{\xx{A}{3}\xx{A}{4}\xx{B}{3}} $\\
     & & \hllime{$+\xI{\xx{A}{1}\xx{A}{2}\xx{A}{3}\xx{B}{1}} +\xI{\xx{A}{1}\xx{A}{2}\xx{A}{3}\xx{B}{2}} +\xI{\xx{A}{1}\xx{A}{3}\xx{B}{1}\xx{B}{2}}$}\\
     & & $ +\xI{\xx{A}{1}\xx{A}{2}\xx{A}{4}\xx{B}{1}}+\xI{\xx{A}{1}\xx{A}{2}\xx{A}{4}\xx{B}{2}}+\xI{\xx{A}{1}\xx{A}{2}\xx{A}{4}\xx{B}{3}} $\\
     & & $ +\xI{\xx{A}{1}\xx{A}{3}\xx{A}{4}\xx{B}{1}}+\xI{\xx{A}{1}\xx{A}{3}\xx{A}{4}\xx{B}{2}}+\xI{\xx{A}{1}\xx{A}{3}\xx{A}{4}\xx{B}{3}} $\\
     & & $ +\xI{\xx{A}{1}\xx{A}{4}\xx{B}{1}\xx{B}{2}}+\xI{\xx{A}{1}\xx{A}{4}\xx{B}{1}\xx{B}{3}}+\xI{\xx{A}{1}\xx{A}{4}\xx{B}{2}\xx{B}{3}} $\\
     & & $ +\xI{\xx{A}{2}\xx{A}{3}\xx{A}{4}\xx{B}{2}}+\xI{\xx{A}{2}\xx{A}{3}\xx{A}{4}\xx{B}{3}}+\xI{\xx{A}{2}\xx{A}{4}\xx{B}{2}\xx{B}{3}} $\\
     & & $ -\xI{\xx{A}{1}\xx{A}{2}\xx{A}{3}\xx{A}{4}\xx{B}{1}}-\xI{\xx{A}{1}\xx{A}{2}\xx{A}{3}\xx{A}{4}\xx{B}{2}}-\xI{\xx{A}{1}\xx{A}{2}\xx{A}{3}\xx{A}{4}\xx{B}{3}} $\\
     & & $ -\xI{\xx{A}{1}\xx{A}{2}\xx{A}{4}\xx{B}{2}\xx{B}{3}}-\xI{\xx{A}{1}\xx{A}{3}\xx{A}{4}\xx{B}{1}\xx{B}{2}}-\xI{\xx{A}{1}\xx{A}{4}\xx{B}{1}\xx{B}{2}\xx{B}{3}} $ \\
    $\QP{5}$ &\Ink{20,45,36,10}  & ... \\
    \hline 
\end{tabular}
\end{center}
\caption{ \ino\ classification and I-basis terms of \qproj\ inequalities.  Recurring terms are highlighted.
}
\label{tab:qp}
\end{table}
Rather than converting to the previous $\x{A,B,C}$ notation, we retain the $\x{A}_i, \x{B}_j$ notation since this is better-suited to comparing terms across different $m$.
In particular, in this notation we can easily identify the recurring terms (indicated by same-color highlight in \cref{tab:qp}); we see that $\QP{n+1}$ includes all terms in $\QP{n}$, suggesting a recursive structure.  By comparing the \ino s to those of the TF building blocks (cf.~\cref{tab:sing}), we extract the general pattern.
In particular, observe that $\QP{2} \sim -I_{(1:1:1)}$, $\QP{3} \sim \QP{2} - 3C_{(1:1:1|1)}$, $\QP{4} \sim \QP{3} - 6 C_{(1:1:1|2)}$, and $\QP{5} \sim \QP{4} - 10 C_{(1:1:1|3)}$. This leads us to conjecture that
\begin{equation} \label{eq:pconj}
    \QP{m} \sim \QP{m-1} - \frac{m(m-1)}{2} C_{(1:1:1|m-2)} \ .
\end{equation}
After a more detailed examination of the structure, we can extract an explicit form for \cref{eq:pconj} as
\begin{equation} \label{eq:conjf}
    \QP{m} = \QP{m-1} - \sum_{i = 1}^{m-1} \sum_{j=i}^{m-1} \fctI{\xs{A}{i}}{\xs{A}{m}}{\xs{B}{j}}{\xm{A}{1}{i-1} \xm{A}{j+1}{m-1-j} \xm{B}{i}{j-i}}.
\end{equation}
Finally, by iteratively applying this recursion formula, we  construct the \ctform\ for the \qproj\ inequalities, summarized by our second key theorem:\footnote{\, 
    Similar to the \qtor\ case, we explicitly present the \ctform\ with the region $\text{B}_{m}$ chosen as the purifier. Nevertheless, owing to the $D_{2m}$ symmetry of the \qproj\ inequality, every element in its permutation-purification orbit admits a \ctform\ of the same structure. In particular, changing the purifier amounts to a permutation of the region labels. The case in which one of the $\text{A}$ regions is chosen as the purifier is discussed explicitly in the following paragraphs and further explained in \cref{ss:projpur}.
}
\begin{thm}\label{thm:projct}
    The \qproj\ family labeled by $m$ is TF-compatible, and takes the form:
    \begin{equation}\label{eq:projct}
        \QP{m} = \sum_{i = 1}^{m-1} \sum_{j = i}^{m-1} - \fctI{\xs{A}{i}}{\xs{B}{j}}{\xm{A}{j+1}{m-j}}{\xm{A}{1}{i-1}\xm{B}{i}{j-i}} \ .
    \end{equation}
\end{thm}
The explicit proof of \cref{eq:projct} is relegated to \cref{ss:thm2pf}. 

%
\begin{table}[htbp] 
\begin{center}
\scriptsize
\begin{tabular}{| c | c | }
\hline
label &  $\QP{m}$ information quantity \\ 
\hline
\hline
$\QP{3}$
    &  \makecell{$-\xS{AB}-\xS{AC}-\xS{AE}-\xS{BC}-\xS{CD}-\xS{DE}-\xS{ABCD}-\xS{ABCE}-\xS{ACDE}$ \\$ +2\xS{ABC}+\xS{ABE}+\xS{ACD}+\xS{ACE}+\xS{ADE}+\xS{BCD}+\xS{CDE}$ }
\\ 
    & \shadeI{
    $+\xI{ABCD}+\xI{ABCE}+\xI{ACDE}-\xI{ABD}-\xI{ACD}-\xI{ACE}-\xI{BCE}$ 
    } 
\\ 
    & \shadeTF{
    $-\tI{A:D:BC}-\ctI{A:E:C}{D}-\ctI{B:E:C}{A}$ 
    } 
\\ \hline
$\QP{4}$
    &   \makecell{
    $-\xS{ABC}-\xS{ABD}-\xS{ABG}-\xS{ACD}-\xS{ADF}-\xS{AFG}-\xS{BCD}-\xS{CDE}-\xS{DEF}-\xS{EFG}-\xS{ABCDE}-\xS{ABCDF} $\\
    $ -\xS{ABCDG}-\xS{ABDFG}-\xS{ACDEF}-\xS{ADEFG}+3\xS{ABCD}+\xS{ABCG}+\xS{ABDF}+\xS{ABDG}+\xS{ABFG}+\xS{ACDE} $\\
    $ +\xS{ACDF}+\xS{ADEF}+\xS{ADFG}+\xS{AEFG}+\xS{BCDE}+\xS{CDEF}+\xS{DEFG}$} 
\\ 
    & \shadeI{
    \makecell{
    $-\xI{ABCDE}-\xI{ABCDF}-\xI{ABCDG}-\xI{ABDFG}-\xI{ACDEF}-\xI{ADEFG}+\xI{ABCE}+\xI{ABCF}+\xI{ABDE}+\xI{ABDF} $\\
    $ +\xI{ABDG}+\xI{ACDE}  +\xI{ACDF} +\xI{ACDG}+\xI{ACEF}+\xI{ADEF}+\xI{ADEG}+\xI{ADFG}+\xI{BCDF}+\xI{BCDG}+\xI{BDFG} $\\
    $ -\xI{ABE}-\xI{ACE}-\xI{ACF}-\xI{ADE}  -\xI{ADF}-\xI{ADG}  -\xI{BCF}-\xI{BDF}-\xI{BDG}-\xI{CDG}$ }
    } 
\\ 
    & \shadeTF{
    $-\tI{A:E:BCD}-\ctI{A:F:CD}{E}-\ctI{A:G:D}{EF}-\ctI{B:F:CD}{A}-\ctI{B:G:D}{AF}-\ctI{C:G:D}{AB}$ 
    } 
\\ \hline
$\QP{5}$
    &   \makecell{
    $-\xS{ABCD}-\xS{ABCE}-\xS{ABCI}-\xS{ABDE}-\xS{ABEH}-\xS{ABHI}-\xS{ACDE}-\xS{ADEG}-\xS{AEGH}-\xS{AGHI}-\xS{BCDE}$\\
    $-\xS{CDEF}-\xS{DEFG}-\xS{EFGH}-\xS{FGHI}-\xS{ABCDEF}-\xS{ABCDEG}-\xS{ABCDEH}-\xS{ABCDEI}-\xS{ABCEHI}-\xS{ABDEGH}$\\
    $-\xS{ABEGHI}-\xS{ACDEFG}-\xS{ADEFGH}-\xS{AEFGHI}
    +4\xS{ABCDE}+\xS{ABCDI}+\xS{ABCEH}+\xS{ABCEI}+\xS{ABCHI}$\\
    $+\xS{ABDEG}+\xS{ABDEH}+\xS{ABEGH}+\xS{ABEHI}+\xS{ABGHI}+\xS{ACDEF}+\xS{ACDEG}+\xS{ADEFG}+\xS{ADEGH}$\\
    $+\xS{AEFGH}+\xS{AEGHI}+\xS{AFGHI}+\xS{BCDEF}+\xS{CDEFG}+\xS{DEFGH}+\xS{EFGHI}$} 
\\ 
    & \shadeI{
    \makecell{
    $+\xI{ABCDEF}+\xI{ABCDEG}+\xI{ABCDEH}+\xI{ABCDEI}+\xI{ABCEHI}+\xI{ABDEGH}+\xI{ABEGHI}+\xI{ACDEFG}+\xI{ADEFGH}$\\
    $+\xI{AEFGHI}-\xI{ABCDF}-\xI{ABCDG}-\xI{ABCDH}-\xI{ABCEF}-\xI{ABCEG}-\xI{ABCEH}-\xI{ABCEI}-\xI{ABDEF}-\xI{ABDEG}$\\
    $-\xI{ABDEH}-\xI{ABDEI}-\xI{ABDGH}-\xI{ABEGH}-\xI{ABEGI}-\xI{ABEHI}-\xI{ACDEF}-\xI{ACDEG}-\xI{ACDEH}-\xI{ACDEI}$\\
    $-\xI{ACDFG}-\xI{ACEFG}-\xI{ACEHI}-\xI{ADEFG}-\xI{ADEFH}-\xI{ADEGH}-\xI{ADFGH}-\xI{AEFGH}-\xI{AEFGI}-\xI{AEFHI}$\\
    $-\xI{AEGHI}-\xI{BCDEG}-\xI{BCDEH}-\xI{BCDEI}-\xI{BCEHI}-\xI{BDEGH}-\xI{BEGHI}+\xI{ABCF}+\xI{ABCG}+\xI{ABDF}$\\
    $+\xI{ABDG}+\xI{ABDH}+\xI{ABEF}+\xI{ABEG}+\xI{ABEH}+\xI{ABEI}+\xI{ACDF}+\xI{ACDG}+\xI{ACDH}+\xI{ACEF}+\xI{ACEG}$\\
    $+\xI{ACEH}+\xI{ACEI}+\xI{ACFG}+\xI{ADEF}+\xI{ADEG}+\xI{ADEH}+\xI{ADEI}+\xI{ADFG}+\xI{ADFH}+\xI{ADGH}+\xI{AEFG}$\\
    $+\xI{AEFH}+\xI{AEFI}+\xI{AEGH}+\xI{AEGI}+\xI{AEHI}+\xI{BCDG}+\xI{BCDH}+\xI{BCEG}+\xI{BCEH}+\xI{BCEI}+\xI{BDEG}$\\
    $+\xI{BDEH}+\xI{BDEI}+\xI{BDGH}+\xI{BEGH}+\xI{BEGI}+\xI{BEHI}+\xI{CDEH}+\xI{CDEI}+\xI{CEHI}$\\
    $-\xI{ABF}-\xI{ACF}-\xI{ACG}-\xI{ADF}-\xI{ADG}-\xI{ADH}-\xI{AEF}-\xI{AEG}-\xI{AEH}-\xI{AEI}-\xI{BCG}-\xI{BDG}-\xI{BDH}$\\
    $-\xI{BEG}-\xI{BEH}-\xI{BEI}-\xI{CDH}-\xI{CEH}-\xI{CEI}-\xI{DEI}$ }
    } 
\\ 
    & \shadeTF{
    \makecell{$-\tI{A:F:BCDE}-\ctI{A:G:CDE}{F}-\ctI{A:H:DE}{FG}-\ctI{A:I:E}{FGH}-\ctI{B:G:CDE}{A}-\ctI{B:H:DE}{AG}$\\
    $-\ctI{B:I:E}{AGH}-\ctI{C:H:DE}{AB}-\ctI{C:I:E}{ABH}-\ctI{D:I:E}{ABC}$ }
    } 
\\ \hline
 \hline 
\end{tabular}
\end{center}
\caption{
Explicit expressions for $\QP{m}$ for small $m$ in the S-basis (white background), I-basis (blue background), and the \ctform\ (pink background). 
The notational shorthand and conventions are the same as in \cref{tab:HEI_toric} and \cref{tab:HEI_N6}.}
\label{tab:HEI_proj}
\end{table}

In \cref{tab:HEI_proj} we display $\QP{m}$ for $m=3,4,5$, but for the sake of compactness as well as easier comparison with the \qtor\ family tabulated in \cref{tab:HEI_toric}, we present the information quantities in the original $\x{A,B,C}$ notation and conventions.
Notice that unlike the \qtor\ family where all terms have unit coefficients, the \qproj\ S-basis expressions have a single entropy term with coefficient $(m-1)$, as manifest in the alternate form of \cref{eq:Qproj2}.  On the other hand, the I-basis form has all terms with unit coefficients, and correspondingly there are no repeated terms in the \ctform.
\begin{table}[htbp]
\begin{center}
    \vTFtab[np, cell size=0.26,box size=0.13, header font=\sffamily\fontsize{5}{12}\selectfont,
    title=$\QP{3}$]{5}{{A, D, BC}, {A, E, C, D}, {B, E, C, A}}
    \qquad
    \vTFtab[np, cell size=0.26,box size=0.13, header font=\sffamily\fontsize{5}{12}\selectfont,
    title=$\QP{4}$]{7}{{A, E, BCD}, {A, F, CD, E}, {A, G, D, EF}, {B, F, CD, A}, {B, G, D, AF}, {C, G, D, AB}}
    \qquad
    \vTFtab[np, cell size=0.26,box size=0.13, header font=\sffamily\fontsize{5}{12}\selectfont,
    title=$\QP{5}$]{9}{{A, F, BCDE}, {A, G, CDE, F}, {A, H, DE, FG}, {A, I, E, FGH}, {B, G, CDE, A}, {B, H, DE, AG}, {B, I, E, AGH}, {C, H, DE, AB}, {C, I, E, \
    ABH}, {D, I, E, ABC}}
    \qquad
    \vTFtab[np, cell size=0.26,box size=0.13, header font=\sffamily\fontsize{5}{12}\selectfont,
    title=$\QP{6}$]{11}{{A, G, BCDEF}, {A, H, CDEF, G}, {A, I, DEF, GH}, {A, J, EF, GHI}, {A, K, F, GHIJ}, {B, H, CDEF, A}, {B, I, DEF, AH}, {B, J, EF, AHI}, {B, K, F, AHIJ}, {C, I, DEF, AB}, {C, J, EF, ABI}, {C, K, F, ABIJ}, {D, J, EF, ABC}, {D, K, F, ABCJ}, {E, K, F, ABCD}}
    \qquad
\end{center}
\caption{Diagrammatic representation of instances of $\QP{m}$ for small $m$.}
\label{tab:QPgraphicalrep}
\end{table}
To further elucidate the TF expressions, \cref{tab:QPgraphicalrep} provides the corresponding diagrammatic representation, now up to $m=6$ (corresponding to $\nN=11$).

As we can see from \cref{tab:HEI_proj} and \cref{tab:QPgraphicalrep}, the complexity of the expressions grows much slower in the S-basis and in the TF than in the I-basis.  To quantify this,  \cref{tab:numSIT4QP} presents the total number of terms of  $\QP{m}$ when written in S-basis, I-basis, and TF.  We will provide a more refined summary in \cref{s:discussion}. 
\begin{table}[htbp]
\begin{center}
\scriptsize
\begin{tabular}{|c||c|c|c|}
    \hline
    & \# S &\# I &\# TF  \\ \hline \hline
    \shadeG{$m=2$} & 7 & \shadeI{1} & \shadeTF{1} \\ 
    \hline
    \shadeG{$m=3$} & 17 & \shadeI{7} & \shadeTF{3} \\ 
    \hline
    \shadeG{$m=4$} & 31 & \shadeI{31} & \shadeTF{6} \\ 
    \hline
    \shadeG{$m=5$} & 49 & \shadeI{111} & \shadeTF{10} \\ 
    \hline
    \shadeG{$m=6$} & 71 & \shadeI{351} & \shadeTF{15} \\ 
    \hline
    \shadeG{$m=7$} & 97 & \shadeI{1023} & \shadeTF{21} \\ 
    \hline
\end{tabular}
\end{center}
\caption{Numbers of terms of $\QP{m}$ written in  the S-basis (white), the I-basis (blue), and the TF of \cref{eq:projct} (pink).}
\label{tab:numSIT4QP}
\end{table}
It is instructive to see how this compares with the \qtor\ family.  Notice that \cref{tab:numSIT4QP} is analogous to \cref{tab:ctterm}, except that the latter tabulates the sizes for a 2-parameter family $\QT{m,n}$ (with odd $m$ and $n$) whereas the former constitutes a 1-parameter family  $\QP{m}$ for any positive $m$.  If we compare the diagonal ($m=n$) terms in $\QT{m,n}$ with the odd $m$ values of $\QP{m}$, we see that the I-basis expressions have the same length for the two families at a given $m$, while the S-basis and TF expressions are slightly longer for the \qtor\ family.  To compare the even-$m$ values of  $\QP{m}$, it seems natural to consider the immediately off-diagonal terms in $\QT{m,n}$ (i.e.\ $m=n+2$ or $n=m+2$, which have the same number of terms in both S-basis and I-basis).  In this case, we find that in fact both S-basis and I-basis numbers match between the two families, but once again, the TF forms of $\QT{m,n}$ are longer (in both off-diagonal terms).  

\begin{figure}
    \centering
        \includegraphics[width=0.9\linewidth]{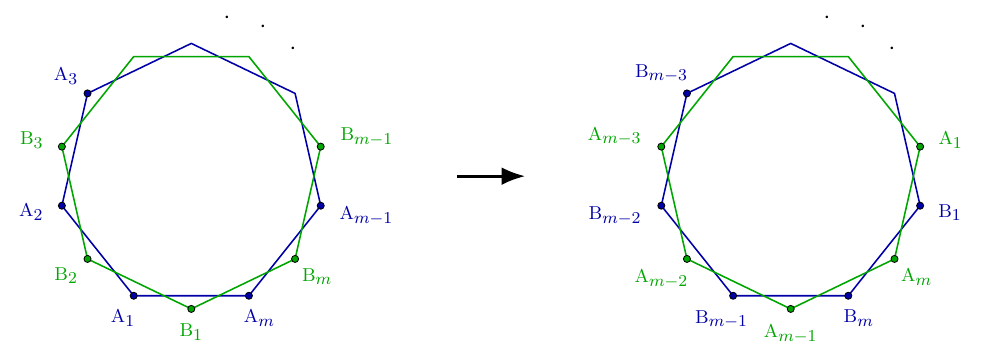}
    \caption{Relabeling of regions in \cref{eq:projpuri}.  
    The $\xs{A}{i}$ and $\xs{B}{i}$ are ordered cyclically (indicated by the blue and green polygons), but they interweave in a way which admits the dihedral $(2m)$ symmetry, which means that our relabeling actually leaves the structure invariant.
    }
    \label{fig:relabel_1}
\end{figure}

This observation justifies our earlier remark that the \qproj\ family has a simpler TF than the \qtor\ family.  However, given the large difference in the latter under swapping $m$ and $n$, or equivalently swapping the $\x{A}_i$ regions with the  $\x{B}_i$ regions, one might wonder if we can't do even better in the \qproj\ case:  could we get a more compact TF expression for $\QP{m}$ if we swap the  $\x{A}_i$ regions with the  $\x{B}_i$ regions?  This is tantamount to changing the purifier.
Performing the corresponding purification, we summarize the transformation in the following lemma:
\begin{lemma}\label{lmm:projpuri}
    The  \qproj\ family labeled by $m$, with $\x{A}_m$ chosen  as the purifier, takes the form:
    \begin{equation}\label{eq:projpuri}
        \QP{m} = \sum_{i = 1}^{m-1} \sum_{j=i}^{m-1} 
        -\fctI{\xs{A}{i}}{\xs{B}{j}}{\xm{B}{m}{i}}{\xm{B}{j+1}{m-j-1}\xm{A}{i+1}{j-i}} \ .
    \end{equation}
\end{lemma}
We will trace through the actual purification in \cref{ss:projpur}; here instead we identify an explicit relabeling of regions which transforms between the two forms.
In particular, starting from \cref{eq:projpuri}, we apply the following simultaneous relabeling of the $A$ and $B$ regions, schematically illustrated in \cref{fig:relabel_1}
\begin{itemize}[nosep]
    \item relabel \seq{\xs{A}{i} \to \xs{B}{m-i}}
    \item relabel \seq{\xs{B}{j} \to \xs{A}{m-j}}
\end{itemize}
Reordering the double sum then yields
\begin{equation}\label{eq:relabel4a}
    \QP{m} = \sum_{i = 1}^{m-1} \sum_{j=i}^{m-1} - \fctI{\xs{B}{m-i}}{\xs{A}{m-j}}{\xm{A}{m-i+1}{i}}{\xm{A}{1}{m-j-1}\xm{B}{m-j}{j-i}} \ ,
\end{equation}
and after defining the summation indices as $i=m-\ell$ and $j=m-k$ and changing the summation order, we get
\begin{equation}\label{eq:relabel4b}
    \QP{m} =\sum_{k = 1}^{m-1} \sum_{l = k}^{m-1} -\fctI{\xs{A}{k}}{\xs{B}{l}}{\xm{A}{l+1}{m-l}}{\xm{A}{1}{k-1}\xm{B}{k}{l-k}} \ .
\end{equation}
This precisely matches \cref{eq:projct}, once we rename $k \to i$ and $\ell \to j$, proving the equivalence.  Note that whereas in \cref{eq:projct} the region $\xs{B}{m}$ is the only region which doesn't appear (and hence $\xs{B}{m}$ corresponds to the purifier), in \cref{eq:projpuri} it is the region $\xs{A}{m}$ which does not appear (and hence corresponds to the purifier).  
In other words, to go from \cref{eq:projct} to \cref{eq:projpuri}, we have performed a purification on $\xs{A}{m}$.  Recall that in general, purifications do not preserve the \ctform\ because they can introduce wrong-sign terms.  In \cref{s:building} we will examine explicitly how these wrong-sign terms cancel out, which elucidates how the purified expression remains TF-compatible.

Notice that unlike in the \qtor\ case where purification changed the structure of the TF substantially, here the structure is actually preserved (because we can relate the two expressions by a simple relabeling).  This is a special feature of the \qproj\ family, and harks back to the dihedral $(2m)$ symmetry, which then doesn't distinguish the $\x{A}$ regions from the $\x{B}$ regions.
Hence the answer to our earlier question is that we cannot obtain a more compact TF expression by purifying.  Indeed, as we explain in \cref{s:discussion}, the TF in \cref{eq:projct} is already maximally compact.

\paragraph{Singleton-tripartite form:}
We close this section by expanding out the \qproj\ TF into the corresponding STF expression, analogously to \cref{eq:QtorSTF}:
\begin{equation}\label{eq:QPSFT}
    \QP{m} = \sum_{k=1}^{m-1} \sum_{i=1}^{k} \sum_{j=i}^{k}
      - \fctI{\xs{A}{i}}{\xs{A}{k+1}}{\xs{B}{j}}{\xm{A}{1}{i-1} \xm{A}{j+1}{k-j} \xm{B}{i}{j-i}} \ .
\end{equation}
The summands have the form $C_{(1:1:1|k-1)}$, and the inner two sums give $(k+1)k/2$ such terms, so that $c_{k-1}=(k+1)k/2$.  The outer sum then collects these into the \cno\  $\Cnk{1,3,6,\ldots}$ with the last non-vanishing entry $c_{m-2}=m(m-1)/2$. Summing these up, the full expression has $ (m^2-1)m/6 $ terms in total. Notice that this precisely matches the number of terms in the toric case for odd $m=n$, namely the STF of $\QT{m,m}$ likewise has $ (m^2-1)m/6 $ terms.  For $m>n$, $\QT{m,n}$ has fewer terms than $\QP{m}$, and conversely, for $m<n$, $\QT{m,n}$ has more terms than $\QP{m}$.  However, both expressions have similar structure involving binomial coefficients.

\section{Purification transmutations}
\label{s:building}

In constructing the \ctform s for the \qcyc\ and \qproj\  families of HEIs presented in \cref{eq:torct} and \cref{eq:projct}, we uncovered several structural features shared by both families. 
Apart from the binomial patterns exhibited by  \cno\ classifications in both cases, the similarity becomes especially apparent when one examines how their respective tripartite forms transform under a change of purifier.  Since these observations may offer useful guidance for identifying more general building blocks of HEIs, in this section we examine the key relations in more detail.
We start in \cref{ss:I3id} by establishing a new lemma, \cref{lmm:equach}, phrased at the level of unconditioned tripartite information sums, which will provide the underlying mechanism in simplifying the purification step in both the \qcyc\ and \qproj\ families. Since the \qproj\ family has a simpler TF structure than the \qcyc\ one, we then proceed in \cref{ss:projpur} to elucidate \cref{lmm:projpuri}, explicitly showing how the wrong-sign terms created by purification manage to cancel out in $\QP{m}$.  Similar mechanism underlies the purification in the \qcyc\ family $\QT{m,1}$, captured by \cref{lmm:QTm1}, which we illustrate in \cref{ss:cycpur}.  There we also briefly comment on the similarities between these structural patterns.

\subsection{New \texorpdfstring{$I_3$}{I3} identity}
\label{ss:I3id}

We have seen already in \cref{eq:multTFs} that we can write multiple TF expressions for a given composite tripartite information.  One may then expect that conversely we can recombine certain sums of conditional tripartite informations in various ways.  To motivate a specific realization of such an identity, recall that the (conditional) multipartite information $I_n$ is invariant under permuting its (first) $n$ arguments.  For example for $n=3$, by equating $I_{3}(\x{A:B:C})=I_{3}(\x{C:A:B})$ and using \cref{eq:I4fromI3}, we obtain
$I_{2}(\x{A:BC}) + I_{2}(\x{B:C}) = I_{2}(\x{C:AB}) + I_{2}(\x{A:B})$, 
and by starting with $n=4$, the above identity would accrete one inert argument in each term.  We can further iterate this relation, to build up more terms with different sequential splits of the two non-inert arguments (now relabeled as $\x{A}_i$), to arrive at the following identity:
\begin{lemma}\label{lmm:equach} 
    For any set of disjoint regions $ \xs{A}{1},\xs{A}{2},\xs{A}{3},\ldots,\xs{A}{m}$, and $\xs{B}{1}$ in $\nN \ge m$ party system,
    \begin{equation}\label{eq:equach}
        \sum_{i = 1}^{m-1}\ftI{\xs{B}{1}}{\xs{A}{i+1}}{\xm{A}{1}{i}} = \sum_{j = 1}^{m-1} \ftI{\xs{B}{1}}{\xs{A}{j}}{\xm{A}{j+1}{m-j}} \ .
    \end{equation}
\end{lemma}

Notice that the two sides of the equation would map to each other under inverting the order of the $A$ regions, in other words under mapping $\pqty{\xs{A}{1} ,\xs{A}{2} ,\ldots, \xs{A}{m}} \to \pqty{\xs{A}{m} ,\xs{A}{m-1} ,\ldots, \xs{A}{1}} $.  (In fact, when  $\nN=m$,  once we purify both sides on $\xs{B}{1}$, the equality holds term-wise at $j=i+1$.)
The identity for general $\nN$ follows from a straightforward rearrangement of conditional tripartite information terms.  Using \cref{eq:TFtoSTFgen}, for each $i$ we obtain
\begin{equation}\label{eq:fl}
    \ftI{\xs{B}{1}}{\xs{A}{i+1}}{\xm{A}{1}{i}} = \sum_{j = 1}^{i} \fctI{\xs{B}{1}}{\xs{A}{i+1}}{\xs{A}{j}}{\xm{A}{j+1}{i-j}} \ .
\end{equation}
Substituting \cref{eq:fl} into the left-hand side of \cref{eq:equach},  exchanging the order of the triangular summation over  $1 \le j \le i \le n$ using 
\begin{equation}\label{eq:sumswitch}
    \sum_{i = 1}^{n}\sum_{j = 1}^{i}  \alpha_{ij}
    = \sum_{j = 1}^{n} \sum_{i = j}^{n} \alpha_{ij} \ ,
\end{equation}
and reapplying another version of \cref{eq:TFtoSTFgen}, we find
\begin{equation}\label{eq:prooflm}
\begin{split}
    & \sum_{i = 1}^{m-1}\ftI{\xs{B}{1}}{\xs{A}{i+1}}{\xm{A}{1}{i}} 
     =
    \sum_{i = 1}^{m-1}\sum_{j = 1}^{i} \fctI{\xs{B}{1}}{\xs{A}{i+1}}{\xs{A}{j}}{\xm{A}{j+1}{i-j}} = \\
    &= \sum_{j = 1}^{m-1} \sum_{i = j}^{m-1} \fctI{\xs{B}{1}}{\xs{A}{i+1}}{\xs{A}{j}}{\xm{A}{j+1}{i-j}}
    = \sum_{j = 1}^{m-1} \ftI{\xs{B}{1}}{\xs{A}{j}}{\xm{A}{j+1}{m-j}}
\end{split}
\end{equation} 
The final expression is precisely the right-hand side of \cref{eq:equach}, completing the proof of \cref{lmm:equach}.
\qed  

To further elucidate the workings of the lemma, let us consider the diagrammatic representation of \cref{eq:prooflm} for $\nN=6$:
\[
    \vTFtab[np]{6}{{F, B, A}, {F, C, AB}, {F, D, ABC}, {F, E, ABCD}}
    =
    \vTFtab[np]{6}{{F, B, A}, {F, C, A, B}, {F, C, B}, {F, D, A, BC}, {F, D, B, C}, {F, D, C}, {F, E, A, BCD}, {F, E, B, CD}, {F, E, C, D}, {F, E, D}}
    =
    \vTFtab[np]{6}{{F, B, A}, {F, C, A, B}, {F, D, A, BC}, {F, E, A, BCD}, {F, C, B}, {F, D, B, C}, {F, E, B, CD}, {F, D, C}, {F, E, C, D}, {F, E, D}}
    =
    \vTFtab[np]{6}{{F, BCDE, A}, {F, CDE, B}, {F, DE, C}, {F, E, D}}
\]
Here the middle equality just amounts to rearranging the terms (i.e.\ rows), the outer two equalities correspond to recombining them, and finally after the last step we can swap the second two arguments on the RHS (i.e.\ colors).

\subsection{Projective family purification}
\label{ss:projpur}

Armed with the above identity, we now turn to \cref{lmm:projpuri}. Starting from \cref{eq:projct}, in which $\xs{B}{m}$ serves as the purifier, we perform a change of purifier to $\xs{A}{m}$.  Rather than implementing this change by relabeling the regions as in \cref{s:proj}, we apply the purification identities directly. 
In each term of \cref{eq:projct}, 
$- \fctI{\xs{A}{i}}{\xs{B}{j}}{\xm{A}{j+1}{m-j}}{\xm{A}{1}{i-1}\xm{B}{i}{j-i}}$,
$\xs{A}{m}$ appears in the third argument. We first use \cref{eq:CI3fromI3} to decompose each conditional tripartite information into two tripartite information terms with opposite signs. The positive term now trades the original $\ldots\xs{A}{m}$ argument with the original conditioning subsystem, while the negative term still contains $\xs{A}{m}$. We then use \cref{eq:I3purif} to replace this argument by the complementary subsystem, thereby removing $\xs{A}{m}$ from all the terms. Summing over the indices yields the following two groups of terms:
\begin{itemize}
    \item plus sign terms:\footnote{\, 
        For $i=j=1$, the third argument of the corresponding term is empty. Here and throughout the remainder of this paper, we adopt the convention that any (conditional) tripartite information vanishes whenever one of its first three arguments is empty; in particular, $\ftI{A}{B}{\emptyset}=0$. We nevertheless retain this vanishing term in the sum for notational uniformity, since excluding it would unnecessarily complicate the summation range.
        }
        \begin{equation}\label{eq:pst}
            + \sum_{i = 1}^{m-1}\sum_{j=i}^{m-1} \ftI{\xs{A}{i}}{\xs{B}{j}}{\xm{A}{1}{i-1}\xm{B}{i}{j-i}}
        \end{equation}
    \item minus sign terms:
        \begin{equation}\label{eq:mst}
            - \sum_{i = 1}^{m-1}\sum_{j = i}^{m-1} \ftI{\xs{A}{i}}{\xs{B}{j}}{\xm{A}{i+1}{j-i}\xm{B}{j+1}{m-j}\xm{B}{1}{i-1}}
        \end{equation}
\end{itemize}
The task is then to show that the wrong sign terms of \cref{eq:pst} get canceled by rewriting of \cref{eq:mst}, so that only minus sign terms remain.

\begin{figure}[htbp]
    \begin{small}
        \begin{center}
            \includegraphics[width=0.95\textwidth]{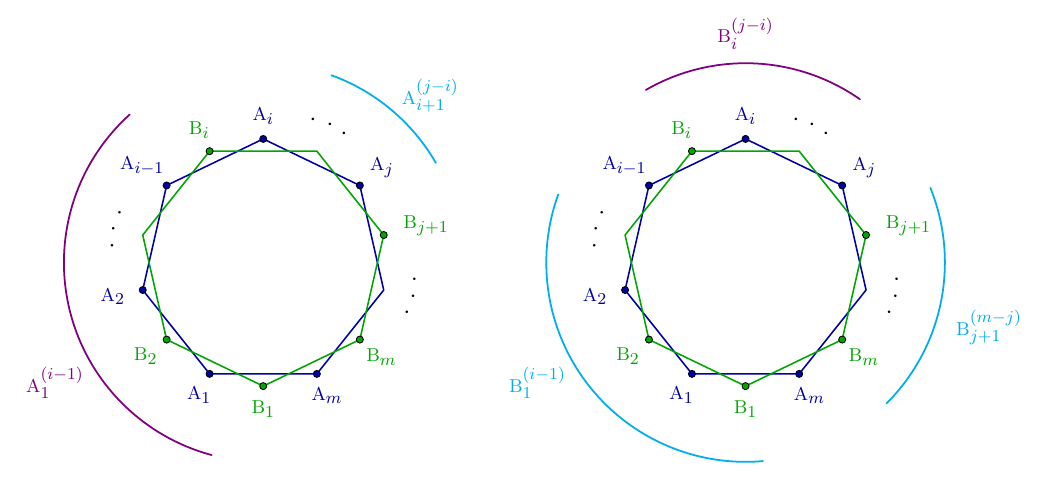}
        \end{center}
        \caption{Regions within the third arguments in \cref{eq:pst,eq:mst}. The left figure displays their $\x{A}$ components, while the right figure displays their $\x{B}$ components. The purple arcs indicate $\xm{A}{1}{i-1}$ and $\xm{B}{i}{j-i}$, which together form the third argument of the plus sign term in \cref{eq:pst}. The teal arcs indicate $\xm{A}{i+1}{j-i}$, $\xm{B}{j+1}{m-j}$, and $\xm{B}{1}{i-1}$, which together form the third argument of the minus sign term in \cref{eq:mst}.}
        \label{fig:newfive}
    \end{small}
\end{figure}

Notice that both groups of terms share the same first two arguments, namely $\xs{A}{i}$ and $\xs{B}{j}$, but differ in their third arguments.  To orient the reader, the extent of these arguments is indicated geometrically by the colored arcs in the polygon representation of the parties shown in \cref{fig:newfive}, color-coded by which sign terms they pertain to (with $A$ and $B$ regions indicated in separate panels for ease of visualization).
We see that they occupy two disjoint portions of the polygon: the third argument of the plus sign term contains the $A$ regions extending from $\xs{A}{1}$ to $\xs{A}{i-1}$, whereas that of the negative-sign term contains a subset of the complementary $\x{A}$ regions extending from $\xs{A}{i+1}$ to $\xs{A}{j}$ and similarly for the $B$ regions. 
This suggests that the expected cancellation of wrong sign terms could be implemented by a relation like \cref{lmm:equach}.  
To see how this works in practice, first decompose the 
 minus sign terms \cref{eq:mst} using \cref{eq:multTFs}:
\begin{equation}\label{eq:transctat}
    \begin{aligned}
        - \sum_{i = 1}^{m-1}\sum_{j = i}^{m-1} 
        \ftI{\xs{A}{i}}{\xs{B}{j}}{\xm{A}{i+1}{j-i}\xm{B}{j+1}{m-j}\xm{B}{1}{i-1}}  
        = & - \sum_{i = 1}^{m-1}\sum_{j = i}^{m-1} 
        \fctI{\xs{A}{i}}{\xs{B}{j}}{\xm{B}{1}{i-1}\xs{B}{m}}{\xm{B}{j+1}{m-j-1}\xm{A}{i+1}{j-i}}\\
         &- \sum_{i = 1}^{m-1}\sum_{j = i}^{m-1} 
         \ftI{\xs{A}{i}}{\xs{B}{j}}{\xm{B}{j+1}{m-j-1}\xm{A}{i+1}{j-i}} \ ,
    \end{aligned}
\end{equation}
and then apply \cref{lmm:equach} to the second line of \cref{eq:transctat} to obtain
\begin{equation}\label{eq:equachrela}
    \sum_{i = 1}^{m-1}\sum_{j = i}^{m-1} \left[ \ftI{\xs{A}{i}}{\xs{B}{j}}{\xm{A}{1}{i-1}\xm{B}{i}{j-i}}  - \ftI{\xs{A}{i}}{\xs{B}{j}}{\xm{B}{j+1}{m-j-1}\xm{A}{i+1}{j-i}} \right] = 0 \ .
\end{equation}
More explicitly, we actually apply \cref{lmm:equach} twice. Starting with the first term in the brackets, we fix $i$ and apply the lemma to the sum over $j$, treating $\xs{B}{i},\xs{B}{i+1},\ldots,\xs{B}{m-1}$ as an ordered sequence. We then interchange the order of summation as in \cref{eq:sumswitch} (with $i\leftrightarrow j$) and apply the lemma again, this time fixing $j$ and summing over $i$. This yields the second term in the brackets.
Finally, we observe that the first term in the brackets is precisely the plus-sign contribution in \cref{eq:pst}, while the second coincides with the second term in \cref{eq:transctat}. After the summation, only the first term in \cref{eq:transctat} remains, which after combining $\xm{B}{1}{i-1}\xs{B}{m}=\xm{B}{m}{i}$, yields exactly the expression in \cref{eq:projpuri}.
\qed

\subsection{Dihedral family purification}
\label{ss:cycpur}

\begin{figure}
    \centering
    \includegraphics[width=0.95\linewidth]{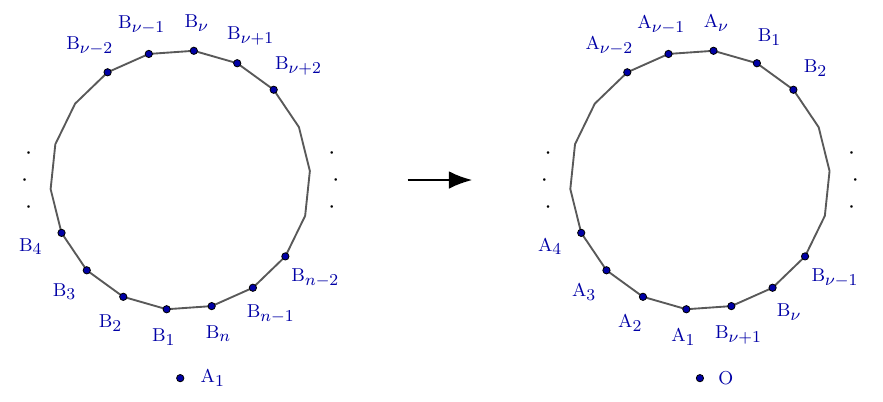}
    \caption{Relabeling of the regions in \cref{eq:Q1n}. The regions are arranged cyclically around a polygon, with one region placed outside. This arrangement makes the $D_n$ symmetry of the \qcyc\ inequality manifest.}
    \label{fig:relabel_2}
\end{figure}

We now turn to the proof of \cref{lmm:QTm1}.  In particular, we will explain how the purification of $\QT{1,n}$ leads to the expression of $\QT{n,1}$ in \cref{eq:Qm1} of \cref{lmm:QTm1}. 
For convenience, we reproduce the starting point \cref{eq:Q1n} below:
\begin{equation*}
    \QT{1,n} = -\sum_{k = 1}^{\nu} \fctI{\xs{A}{1}}{\xs{B}{\nu+k}}{\xm{B}{k}{\nu-k+1}}{\xm{B}{\nu+1}{k-1}}
\end{equation*}
Guided by the preceding discussion of the \qproj\ inequalities, we first relabel the regions to bring this expression into an analogous form. One convenient choice, illustrated in \cref{fig:relabel_2}, is the following:
\begin{itemize}[nosep]
    \item relabel $\xs{A}{1} \to \x{O}$
    \item for $k \in \{1, \cdots , \nu+1 \}$, relabel $\xs{B}{\nu+k} \to \xs{B}{k}$
    \item for $k \in \{1, \cdots , \nu \}$, relabel $\xs{B}{k} \to \xs{A}{k}$
\end{itemize}

Under this relabeling, and renaming the summation index $k \to j$, the expression becomes
$$
 \QT{1,n} = -\sum_{j = 1}^{\nu}  \fctI{\x{O}}{\xs{B}{j}}{\xm{A}{j}{\nu-j+1}}{\xm{B}{1}{j-1}}
$$
and rendering this in STF using \cref{eq:TFtoSTFgen} (and reordering the arguments), we obtain
\begin{equation}\label{eq:Q1nre}
    \QT{1,n} = -\sum_{j = 1}^{\nu}\sum_{i = j}^{\nu} \fctI{\xs{A}{i}}{\xs{B}{j}}{\x{O}}{\xm{A}{i+1}{\nu-i}\xm{B}{1}{j-1}} \ . 
\end{equation}
Notice that $\xs{B}{\nu+1}$ is absent from every term (thereby acting as the purifier) while $\x{O}$ is present in every term.  So the next step is to change the purifier from $\xs{B}{\nu+1}$ to $\x{O}$ (so that later, we can simply rename the $B$ regions into the remaining $A$ regions, to obtain the expression in terms of $A$ regions only, as desired for \cref{eq:Qm1}.)
To implement this transformation, we proceed as in the \qproj\ case. Using \cref{eq:CI3fromI3}, we decompose each conditional tripartite information into two tripartite information terms with opposite signs. The positive term has the original conditioning subsystem, $\xm{A}{i+1}{\nu-i}\xm{B}{1}{j-1}$, as its third argument. The third argument of the negative term still contains $\x{O}$, so we use \cref{eq:I3purif} to replace it by its complement. Summing over $i$ and $j$ therefore produces the following two groups of terms:
\begin{itemize}
    \item plus sign terms:
        \begin{equation}\label{eq:pst2}
            + \sum_{j = 1}^{\nu}\sum_{i = j}^{\nu} \ftI{\xs{A}{i}}{\xs{B}{j}}{\xm{B}{1}{j-1}\xm{A}{i+1}{\nu-i}} 
        \end{equation}
    \item minus sign terms:
        \begin{equation}\label{eq:mst2}
            - \sum_{j = 1}^{\nu}\sum_{i = j}^{\nu} \ftI{\xs{A}{i}}{\xs{B}{j}}{\xm{A}{1}{i-1}\xm{B}{j+1}{\nu-j+1}} 
        \end{equation}
\end{itemize}
As previously, we can split the plus sign terms into two sets.  For each fixed $j$, we apply \cref{lmm:equach} to the ordered sequence $\xs{A}{j},...,\xs{A}{\nu},\xm{B}{1}{j-1}$, treating $\xm{B}{1}{j-1}$ as a single region and taking $\xs{B}{j}$ to play the role of $\xs{B}{1}$ in the lemma, which gives
\begin{equation}
    \begin{split}
            + \sum_{j = 1}^{\nu}\sum_{i = j}^{\nu} \ftI{\xs{A}{i}}{\xs{B}{j}}{\xm{B}{1}{j-1}\xm{A}{i+1}{\nu-i}} = & +\sum_{j=1}^{\nu-1}\sum_{i=j}^{\nu-1} \ftI{\xs{A}{i+1}}{\xs{B}{j}}{\xm{A}{j}{i-j+1}}\\
    & + \sum_{j = 1}^{\nu} \ftI{\xs{B}{j}}{\xm{A}{j}{\nu-j+1}}{\xm{B}{1}{j-1}} \ .
    \end{split}
\end{equation}
Similarly the minus sign terms decompose as
\begin{equation}
    \begin{split}
         - \sum_{j = 1}^{\nu}\sum_{i = j}^{\nu} \ftI{\xs{A}{i}}{\xs{B}{j}}{\xm{A}{1}{i-1}\xm{B}{j+1}{\nu-j+1}}  =& -\sum_{j = 1}^{\nu-1}\sum_{i=j}^{\nu-1}\ftI{\xs{A}{i+1}}{\xs{B}{j}}{\xm{A}{j}{i-j+1}}\\
    &  - \sum_{j=1}^{\nu}\ftI{\xs{B}{j}}{\xm{A}{j}{\nu-j+1}}{\xm{A}{1}{j-1}\xm{B}{j+1}{\nu-j+1}} \ .
    \end{split}
\end{equation}

Adding them together, the first term cancels, resulting in:
\begin{equation}\label{eq:totalpuri}
    \begin{split}
        \QT{1,n} = & - \sum_{j=1}^{\nu}\ftI{\xs{B}{j}}{\xm{A}{j}{\nu-j+1}}{\xm{A}{1}{j-1}\xm{B}{j+1}{\nu-j+1}} + \sum_{j = 1}^{\nu} \ftI{\xs{B}{j}}{\xm{A}{j}{\nu-j+1}}{\xm{B}{1}{j-1}}\\
         = & - \sum_{j = 2}^{\nu} \fctI{\xs{B}{j}}{\xm{A}{j}{\nu-j+1}}{\xm{A}{1}{j-1}}{\xm{B}{j+1}{\nu-j+1}}\\
         & - \sum_{j=1}^{\nu}\ftI{\xs{B}{j}}{\xm{A}{j}{\nu-j+1}}{\xm{B}{j+1}{\nu-j+1}} + \sum_{j = 1}^{\nu} \ftI{\xs{B}{j}}{\xm{A}{j}{\nu-j+1}}{\xm{B}{1}{j-1}}
    \end{split}
\end{equation}
Applying \cref{lmm:equach} and using \cref{eq:CI3fromI3} to combine the resulting differences into conditional tripartite information terms, we can rewrite the last line as
\begin{equation}\label{eq:middlecyc}
    \begin{split}
        &- \sum_{j=1}^{\nu}\ftI{\xs{B}{j}}{\xm{A}{j}{\nu-j+1}}{\xm{B}{j+1}{\nu-j+1}}+ \sum_{j = 1}^{\nu} \ftI{\xs{B}{j}}{\xm{A}{j}{\nu-j+1}}{\xm{B}{1}{j-1}} \\ 
       = &- \sum_{i = 1}^{\nu}\sum_{j = 1}^{i}\fctI{\xs{B}{j}}{\xs{A}{i}}{\xm{B}{i+1}{\nu-i+1}}{\xm{A}{i+1}{\nu-i}\xm{B}{j+1}{i-j}} \ .
    \end{split}
\end{equation}
To see this more explicitly, we focus on the LHS. We first iterate \cref{eq:CI3fromI3} to single out each region $\xs{A}{i}$ from the composite argument $\xm{A}{j}{\nu-j+1}$, while placing the remaining regions $\xm{A}{i+1}{\nu-i}$ in the conditioning argument, as in our construction of the STFs \cref{eq:TFtoSTFgen}. This introduces an additional sum over $i$, after which we exchange the order of summation. For each fixed $i$, we apply \cref{lmm:equach} to the ordered sequence $\xs{B}{1},\xs{B}{2},\ldots,\xs{B}{i}$, with $\xs{A}{i}$ fixed and $\xm{A}{i+1}{\nu-i}$ carried along as a common conditioning subsystem. The lemma then relates the sum containing the subsystems $\xm{B}{1}{j-1}$ to the corresponding sum containing the subsystems $\xm{B}{j+1}{i-j}$. We next split the original third argument of the first term on the LHS of \cref{eq:middlecyc} as
\begin{equation}
    \xm{B}{j+1}{\nu-j+1}
    =
    \xm{B}{j+1}{i-j}\xm{B}{i+1}{\nu-i+1}.
\end{equation}
Applying \cref{eq:multTFs} to this decomposition produces one group of terms in which $\xm{B}{j+1}{i-j}$ remains a main argument and another in which it is included in the conditioning subsystem. The former cancels against the terms of the second part of LHS due to \cref{lmm:equach}, while the latter gives precisely the conditional tripartite information terms on the RHS.

\begin{figure}
    \centering
    \includegraphics[width=0.95\linewidth]{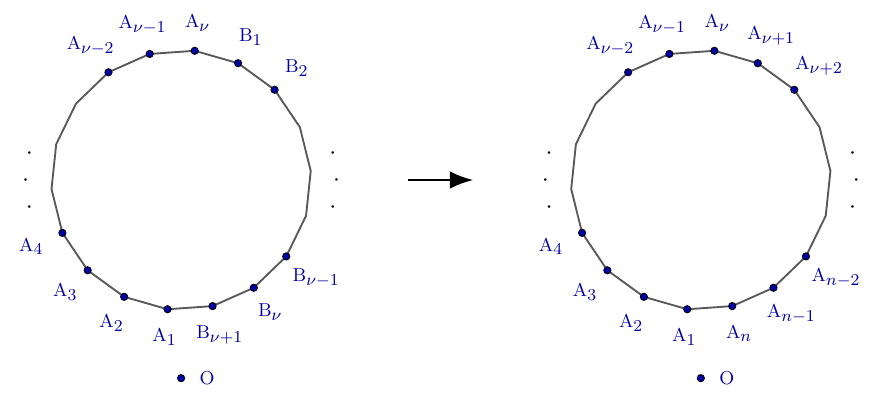}
    \caption{Relabeling of the regions in \cref{eq:lmm3p2}.}
    \label{fig:relabel3}
\end{figure}

Substituting this identity into \cref{eq:totalpuri}, the remaining terms combine into the desired tripartite form:
\begin{equation}\label{eq:lmm3p2}
    \begin{split}
        \QT{1,n} = & - \sum_{i = 1}^{\nu}\sum_{j = 1}^{i}\fctI{\xs{B}{j}}{\xs{A}{i}}{\xm{B}{i+1}{\nu-i+1}}{\xm{A}{i+1}{\nu-i}\xm{B}{j+1}{i-j}}\\
    &- \sum_{j = 2}^{\nu} \fctI{\xs{B}{j}}{\xm{A}{j}{\nu-j+1}}{\xm{A}{1}{j-1}}{\xm{B}{j+1}{\nu-j+1}}
    \end{split}
\end{equation}
We then perform another relabeling of the regions back to the canonical form of $\QT{n,1}$, illustrated as in \cref{fig:relabel3}:
\begin{itemize}[nosep]
    \item for $k \in \{1, \cdots , \nu+1 \}$, relabel $\xs{B}{k} \to \xs{A}{k+\nu}$
    \item for $k \in \{1, \cdots , \nu \}$, leave $\xs{A}{k} \to \xs{A}{k}$
\end{itemize}
Finally, after relabeling $\nu\to\mu$ this leads us exactly to the result (\cref{eq:Qm1}) in \cref{lmm:QTm1}.
\qed

\paragraph{Similarities between projective and \qcyc\ family:}

The proof of \cref{lmm:QTm1} is inspired by a relabeling of the $(m,1)$ \qtor\ family to reveal a structural similarity with the \qproj\ family. We recall the corresponding expressions explicitly below.
\begin{itemize}
    \item \qproj\ family, cf.~\cref{eq:projct}\footnote{\, 
        This expression is obtained from \cref{eq:projct} by exchanging the $\x{A}$ and $\x{B}$ regions and interchanging the summation indices $i$ and $j$.} 
         \begin{equation}
            \QP{m} =- \sum_{j = 1}^{m-1} \sum_{i = j}^{m-1} \fctI{\xs{A}{i}}{\xs{B}{j}}{\xm{B}{i+1}{m-i}}{\xm{A}{j}{i-j}\xm{B}{1}{j-1}}
        \end{equation}

    \item \qtor\ family, cf.~\cref{eq:Q1nre}
        \begin{equation}
            \QT{1,n} = -\sum_{j = 1}^{\nu}\sum_{i = j}^{\nu} \fctI{\xs{A}{i}}{\xs{B}{j}}{\x{O}}{\xm{A}{i+1}{\nu-i}\xm{B}{1}{j-1}}
        \end{equation}
\end{itemize}
Heuristically, both of the HEIs take a form of:
\begin{equation}\label{eq:simform}
    Q = - \sum_{j = 1}^{r} \sum_{i=j}^{r} \fctI{\xs{A}{i}}{\xs{B}{j}}{\x{Purifier Term}}{\x{A,B Combination Term}}
\end{equation}
Although this comparison alone does not provide a general characterization, the structural resemblance between the two infinite families may offer useful guidance for investigating information quantities that remain TF-compatible throughout their permutation-purification orbits. More broadly, comparing the tripartite forms of known families may help identify structural features relevant to full orbit TF compatibility and, ultimately, suggest more general building blocks for constructing such quantities.

\section{Discussion}
\label{s:discussion}

In this work we have examined the two infinite families of HEIs presented in \cite{Czech:2023xed}, namely the toric family of \cref{eq:Qtoric} and the projective family of \cref{eq:Qproj}.  We demonstrated that \emph{all} of these information quantities are TF-compatible, with their tripartite forms given explicitly in the two key theorems, \cref{eq:torct} and \cref{eq:projct}.  This means that \emph{all hitherto-discovered superbalanced HEIs are TF-compatible}, thereby providing highly non-trivial evidence for \cref{conj:HEIsTF}.  
Indeed, if we could actually prove \cref{conj:HEIsTF}, this would give us valuable insights into the structure of the holographic entropy cone, a quest which has now lasted over a decade, formulated with the aim of elucidating spacetime emergence in holography.
However even in the absence of the proof's structural insights, the validity of \cref{conj:HEIsTF} has a number of useful applications and consequences.  Before discussing these in full generality, we first consider what advantages the TF recasting bears for the  toric family of \cref{eq:Qtoric} and the projective family of \cref{eq:Qproj}. 

\paragraph{Compactness of TF:}
As emphasized in the Introduction, the toric and projective families have been originally introduced in the S-basis, utilizing the symmetries manifest in this rendition.  Unlike most of the $\nN=6$ HEIs whose I-basis expression is more compact than the S-basis one, for $\QT{m,n}$ and $\QP{m}$ the I-basis expression is far lengthier.  To get a more detailed sense of the distribution of  the number of $\sI_k$ terms over $k$, 
\cref{tab:incT} presents the \ino\ classification for $\QT{m,n}$ for the first few odd values of $m$ and $n$, from which we can also read off  the \ino\ for $\QP{m}$ as indicated in the caption.
\begin{table}[htbp]
\begin{center}
\scriptsize
\begin{tabular}{|c||c|c|c|c|}
    \hline 
    \shadeG{} & \shadeG{$n = 1$} & \shadeG{$n = 3$} & \shadeG{$n = 5$} & \shadeG{$n = 7$}  \\ \hline\hline
    \shadeG{$m = 1$} & \shadeO{} &\shadeT{\Ink{1}} & \Ink{3, 2}& \Ink{6, 8, 3} \\ \hline
    \shadeG{$m = 3$} & \shadeT{\Ink{1}}& \shadeO{\Ink{4, 3}} & \shadeT{\Ink{10, 15, 6} }& \Ink{19, 42, 33, 9} \\ \hline
    \shadeG{$m = 5$} & \Ink{5,  5,  1} & \shadeT{\Ink{10, 15, 6}} & \shadeO{\Ink{20, 45, 36, 10}} &  \shadeT{\Ink{35, 105, 126, 70, 15}}\\ \hline
    \shadeG{$m = 7$} &\Ink{14,  28,  21,  7,  1}& \Ink{21, 49, 42, 14, 1} & \shadeT{\Ink{35, 105, 126, 70, 15}} & \shadeO{\Ink{56, 210, 336, 280, 120, 21}} \\ 
    \hline
\end{tabular}
\end{center}
\caption{The i\# classification of $\QT{m,n}$ for the specified odd $m$ and $n$.  The diagonal entries (shaded orange) simultaneously correspond to $\QP{m}$, while the entries immediately adjacent to the diagonal with  $\lvert m - n \rvert = 2$ (shaded teal and symmetric under the exchange $m \leftrightarrow n$) simultaneously correspond to $\QP{(m+n)/2}$.} 
\label{tab:incT}
\end{table}
We see that \ino s grow quite rapidly with $n$ and especially $m$.  Since the TF is constructed from the tripartite and conditional tripartite terms, one might expect such large \ino s would lead to very complicated looking TFs.  The remarkable result of our explorations is that in fact for both families, the TF is actually quite simple, and particularly so for $\QP{m}$.
\begin{table}[htbp]
\begin{center}
\scriptsize
\begin{tabular}{|c||c|c||c|c||c|c||c|c|}
\hline 
\shadeG{} & \#S$_{_\text{\color{QPcolor}{P}}}$ & \#S$_{_\text{\color{QTcolor}{T}}}$ & \ino & \shadeI{\#I }& \cno & \#STF & \shadeTF{\#TF$_{_\text{\color{QPcolor}{P}}}$} & \shadeTF{\#TF$_{_\text{\color{QTcolor}{T}}}$}
    \\ \hline\hline
\shadeG{$m = 2$} & 7 & 7
    &\Ink{1} & \shadeI{1} &
    \Cnk{1} & 1 & \shadeTF{1} & \shadeTF{1}
    \\ \hline
\shadeG{$m = 3$} & 17 & 19
    &\Ink{4, 3} & \shadeI{7} &
    \Cnk{1, 3} & 4 & \shadeTF{3} & \shadeTF{4}
    \\ \hline
\shadeG{$m = 4$} & 31 & 31
    &\Ink{10, 15, 6} & \shadeI{31} &
    \Cnk{1, 3, 6} & 10 & \shadeTF{6} & \shadeTF{7}
    \\ \hline
\shadeG{$m = 5$} & 49 & 51
    &\Ink{20, 45, 36, 10} & \shadeI{111} &
    \Cnk{1, 3, 6, 10} & 20 & \shadeTF{10} & \shadeTF{14}
    \\ \hline
\shadeG{$m = 6$} & 71 & 71
    &\Ink{35, 105, 126, 70, 15} & \shadeI{351} &
    \Cnk{1, 3, 6, 10, 15} & 35 & \shadeTF{15} & \shadeTF{19}
    \\ \hline
\shadeG{$m = 7$} & 97 & 99
    &\Ink{56, 210, 336, 280, 120, 21} & \shadeI{1023} &
    \Cnk{1, 3, 6, 10, 15, 21} & 56 & \shadeTF{21} & \shadeTF{29}
    \\ \hline
\end{tabular}
\end{center}
\caption{The number of terms in various expressions for $\QP{m}$, and $\QT{m,m}$ if $m$ is odd, or $\QT{m-1,m+1}$ if $m$ is even.   For the \ino\ and \cno\ classifications, and therefore for the total \#\ of terms in the I-basis and in STF, the quantities coincide between the projective and the toric families, whereas the total \# of terms in the S-basis and in TF renditions differ slightly between the two families. The respective columns are each split into the projective and toric families, as indicated by the subscript in the heading.}
\label{tab:numsum}
\end{table}
To summarize the relevant numerology, \cref{tab:numsum} expands on \cref{tab:incT} to consolidate the number of terms required by the various forms for $\QP{m}$ and (near-)diagonal cross-section of $\QT{m,n}$.  Part of this was presented separately in \cref{tab:ctterm} and \cref{tab:numSIT4QP}, which show the total number of terms in S-basis, I-basis, and TF expressions for $\QT{m,n}$ and $\QP{m}$, respectively.
In \cref{tab:numsum} we additionally include the \ino\ and \cno\ classifications, along with the total number of terms in STF, which all coincide between the two families (where for the toric representatives this equivalence holds for the near-diagonal terms, $\QT{m,m}$ if $m$ is odd, or $\QT{m-1,m+1}$ if $m$ is even).

From \cref{tab:numsum} we see several interesting features: 
First, the \cno\ classifications build incrementally with $m$: each successive value adds only the component $c_{m-2}$, leaving all preceding components unchanged, as already follows from \cref{eq:pconj}. Since the last nonzero entries of the \cno\ and \ino\ classifications agree, this sequence can equivalently be obtained by collecting the last entry of each \ino\ classification as $m$ increases. 
Second, the total number of TF terms in our expression for $\QP{m}$ equals $i_{m+1}$, the number of highest-order I-basis terms. Since each term in any TF decomposition of $\QP{m}$ can contribute at most one $\sI_{m+1}$, this equality shows that our expression is maximally compact: each $\sI_{m+1}$ is grouped with lower-order $\sI_k$'s into a single TF term, with no additional TF terms required. In contrast, our TF expressions for the toric representatives with $m>2$ in \cref{tab:numsum} contain more than $i_{m+1}$ terms. Thus, despite the identical \ino\ and \cno\ classifications and the nearly identical S-basis term counts, these toric TF expressions group the I-basis terms less compactly than their projective counterparts.

Finally, at large $m$ and $n$ the number of terms in the various representations scales as indicated in \cref{tab:scaling}.
\begin{table}[htbp]
\begin{center}
\scriptsize
\begin{tabular}{|c|c|c|c|}
    \hline 
    \shadeG{family} & \# S  & \shadeI{\# I } & \shadeTF{\# TF} \\
    \hline\hline
    \shadeG{$\QT{m,n}$} & $\mathcal{O}(2mn)$ & \shadeI{$\mathcal{O}(2^m + mn \, 2^{(m+n)/2})$} & \shadeTF{$\mathcal{O}(m^2/8+mn/2)$} \\
    \hline
    \shadeG{$\QP{m}$} & $\mathcal{O}(2m^2)$ & \shadeI{$\mathcal{O}(m^2 \, 2^{m})$} & \shadeTF{$\mathcal{O}(m^2/2)$} \\
    \hline
\end{tabular}
\end{center}
\caption{Asymptotic scaling of the number of terms in the various representations for the toric and projective families.
}
\label{tab:scaling}
\end{table}
We see that the number of I-basis terms grows exponentially faster (by a factor of $2^m$ when $m \sim n \gg 1$) than that of S-basis and TF.  Near this diagonal or if $n>m$, the TF rendition is the most compact one: the number of TF terms is smaller by a factor of $\sim 4$ than in the S-basis, though when $m>12n$ it exceeds the S-basis count.

Besides the high compactness of the TF packaging, the actual form of the TF expressions is interesting in itself.  As  demonstrated in the previous section, it is highly constrained, manifesting a rigid structure with nice transformation properties under purifications, and interesting similarity between the two families.

\paragraph{Utility of TF:}
To exemplify the uses of TF, we make several observations about the structural properties of (primitive superbalanced) HEIs.  Many of these properties have originally been guessed based on empirical evidence, but they in fact follow directly from the tripartite form, typically by simply checking the conjectured property in a single term $-\xctI{X:Y:Z}{W} $ which often becomes most manifest in its S-basis representation
\begin{equation}\label{eq:CI3inSbasis}
\begin{split}
    -\xctI{X:Y:Z}{W} = &+ S(\x{XYW}) + S(\x{XZW}) + S(\x{YZW}) + S(\x{W}) \\
        &- S(\x{XYZW}) - S(\x{XW}) - S(\x{YW}) - S(\x{ZW}) \ , 
\end{split}
\end{equation}
and then using the sign definiteness of the TF to argue the said property  prevails in any TF.   To formulate these properties, it will be convenient to recast the HEI into a more conventional form $\text{LHS} \ge \text{RHS}$ in S-basis.  Since any primitive HEI has rational coefficients, we can always express it (by a suitable rescaling if necessary) in a (maximally reduced) form
\begin{equation}\label{eq:HEI_LHS_RHS}
    \sum_{\ell=1}^{m_{_\text{L}}} S(\x{X}_\ell)
    \ge
    \sum_{r=1}^{m_{_\text{R}}} S(\x{Y}_r)
\end{equation}
where the $\x{X}_\ell$ and $\x{Y}_r$ correspond to subsystems of $[\nN]$ (which can be repeated on either side but are distinct between the LHS and RHS), and where all terms appear with +1 coefficient.

\begin{itemize} 

\item Number of terms on two sides of HEI:\\
For all superbalanced HEIs, we empirically observe that the number of terms on the LHS is (strictly) smaller than the  number of terms on the RHS.  The fact that $m_{_\text{L}} \le m_{_\text{R}}$ follows immediately from a TF form of $Q$ by applying the above strategy.  In particular, consider the single term $-\xctI{X:Y:Z}{W}$ recast in the form \cref{eq:HEI_LHS_RHS}.
As manifest from \cref{eq:CI3inSbasis}), if $\x{W}\ne \emptyset$, then both LHS and RHS have 4 terms, whereas if $\x{W}=\emptyset$, LHS now has only 3 terms while RHS still has 4 terms.
Summing over all TF terms (which cannot cancel between separate terms on the same side since all  coefficients in  \cref{eq:HEI_LHS_RHS} are positive), we immediately arrive at $m_{_\text{L}} \le m_{_\text{R}}$.\footnote{\,
A similar term counting argument appears in \cite{Bao:2024azn,Naskar:2024mzi}, where it is used to derive further results.
}  Moreover, since any HEI is sign-definite, its TF should\footnote{\, \label{fn:HEIhasIterms}
    While we are unaware of a rigorous proof, we have strong empirical evidence \cite{Hernandez-Cuenca:2023iqh,Grimaldi:2025jad} (see also \cite{Grimaldi:2026lbq}), bolstered by the expectation that, since each conditioned tripartite information term by itself is not  sign-definite (in fact we can easily find configurations \cite{Hernandez-Cuenca:2023iqh} on which the corresponding $-\xctI{X:Y:Z}{W}<0$), any combination of conditioned tripartite information terms will likewise admit a violating configuration. 
} contain at least one unconditioned $-I_3$ term, which then implies the stronger statement $m_{_\text{L}} < m_{_\text{R}}$.
More specifically, the difference $m_{_\text{R}}-m_{_\text{L}}$ corresponds to the number of unconditioned $I_3$ terms in any TF.

This result immediately demonstrates that \emph{all} primitive superbalanced HEIs can be violated by physically admissible quantum states: a simple example of a violating state (first used in this context for MMI in \cite{Hayden:2011ag}) is the GHZ state on $\nN+1$ parties, since in this state all subsystem entropies are equal.

\item Subsystem inclusion:  \\
For an inequality \eqref{eq:HEI_LHS_RHS} to be a valid HEI, the terms on the two sides cannot be arbitrary.  One simple necessary condition is that for each subsystem $\x{X}_\ell$ on the LHS, there has to be some larger subsystem $\x{Y}_r$ on the RHS  containing it.\footnote{\,
    See also a recent discussion in \cite{Czech:2026tgj} which interpreted this property in terms of RG flow.
} 
The statement that $\forall \ \x{X}_\ell, \ \exists \ \x{Y}_r \supset \x{X}_\ell$ is indeed manifest from  \cref{eq:CI3inSbasis}. 
In fact it is subsumed by a stronger property coined in \cite{Grimaldi:2026lbq} called \emph{region dominance}, which states that for any ``region'' or subsystem $\x{Z}$, the number of LHS terms containing it can be no larger than the number of RHS terms containing it, 
$\abs{\{\x{X}_\ell \mid \x{X}_\ell \supseteq \x{Z}\}} \le \abs{\{\x{Y}_r \mid \x{Y}_r \supseteq \x{Z}\}}$.
Region dominance was proved for any balanced HEI in \cite[Thm.5]{Grimaldi:2026lbq}, but using TF the result follows directly from \cref{eq:CI3inSbasis}, where we can explicitly check that region dominance holds for each TF term individually.

More interestingly, superbalanced HEIs also have the opposite inclusion property, that for each subsystem $\x{X}_\ell$ on the LHS, there has to be some smaller subsystem $\x{Y}_r$ on the RHS which is contained in it.\footnote{\,  
    This is not directly  manifest within \cref{eq:CI3inSbasis}, because it fails on the $\x{W}$ term which comes with positive coefficient in $-\xctI{X:Y:Z}{W}$; though it holds on all the other terms.  However, empirically, it is usually easy to find some  TF representation wherein $-\xctI{X:Y:Z}{W}$ is accompanied by another term where $\x{W}$ is contained in one of the main arguments, so that it gets canceled from the LHS.
}
To see this, recall that HEIs remain HEIs under purifications.  Given an HEI, consider any LHS term $\x{X}_\ell$ and pick any party contained in $\x{X}_\ell$; for definiteness we'll call this party $\x{A}$.  Purify the HEI on $\x{A}$, so that $\x{X}_\ell$ becomes $\x{X}'_\ell = \eval{\overline{\x{X}_\ell}}_{\text{O}\to\text{A}}$ in the new HEI (in other words, we swap the term for its complement and then rename the purifier back to $\x{A}$).  
TF-compatibility implies that the purified HEI is likewise TF-compatible, so by the previous inclusion argument, $\x{X}'_\ell$ must be contained in some larger RHS term $\x{Y}'_r$ which therefore also contains $\x{A}$.  Now purify once again on $\x{A}$, which gets us back to the original HEI.  Then we recover the original term $\eval{\overline{\x{X}'_\ell}}_{\text{O}\to\text{A}} = \x{X}_\ell$, and now we have identified a RHS term $\x{Y}_r=\eval{\overline{\x{Y}'_r}}_{\text{O}\to\text{A}} $ which must be contained in $\x{X}_\ell$, since
$$
    \x{X}'_\ell \subset \x{Y}'_r \qquad \Longleftrightarrow \qquad
    \x{X}_\ell \supset \x{Y}_r \ .
$$

\item Majorization:\\
In fact both inclusion properties follow from the results of \cite{Grimaldi:2025jad}, which we first briefly contextualize here.
Motivated by the question of whether allowing for time-dependent geometric states can exhibit richer entanglement structures than the static case (alternatively phrased as the question of whether or not the ``HRT cone is the RT cone''), the paper \cite{Grimaldi:2025jad} explored configurations which saturate HEIs nontrivially, to see whether these could lead to violations under time-dependent perturbations.  The effect of such perturbations on a given HEI can be analyzed by considering its so-called null reduction, and its positivity can be phrased as a certain majorization property  \cite{Grimaldi:2025jad}, which effectively amounts to the RHS subsystems being more dispersed than the LHS ones. The statements that if a superbalanced $Q$ corresponds to a valid HEI, its null reduction majorizes, and that it remains a valid (non-primitive) HEI, were both conjectured in \cite{Grimaldi:2025jad} with strong evidence based on all known $\nN=6$ HEIs of \cite{Hernandez-Cuenca:2023iqh}, and later proved in \cite{Grimaldi:2026lbq} (without using TF).  This suffices to show the above assertion that   $\forall \ \x{X}_\ell, \ \exists \ \x{Y}_{r_-} , \x{Y}_{r_+}$ such that $\x{Y}_{r_-} \subset \x{X}_\ell \subset \x{Y}_{r_+}$.
The utility of TF-compatibility here was exemplified already by the proof \cite[Thm.2]{Grimaldi:2025jad} to the positivity conjecture which preceded \cite{Grimaldi:2026lbq}.

\item Subsystem intersection:\\
One may also observe that the intersection of the subsystems of each side on the HEIs is empty, i.e.\ $\cap_\ell \x{X}_\ell = \cap_r \x{Y}_r = \emptyset$.
By the above term inclusion property ($\forall \ \x{X}_\ell , \ \exists \ \x{Y}_r \subset \x{X}_\ell$), it suffices to show that $\cap_\ell \x{X}_\ell= \emptyset$ (which was in fact also noted as a corollary in \cite{Czech:2026zca}).  Consider again a single term in TF, $-\xctI{X:Y:Z}{W}$.  Then by \cref{eq:CI3inSbasis}, $\cap_\ell \x{X}_\ell = \x{W}$.  If $\x{W} = \emptyset$ on any term in the TF (as expected from positivity, cf.~\cref{fn:HEIhasIterms}), we are done.  Otherwise, for the intersection to \emph{not} be empty, we would need \emph{each} term $\x{W}_i$ in the TF \cref{eq:ctform} to contain the same party, say $\cap_i \x{W}_i =\x{A}$.  Then we can easily find configurations\footnote{\, 
    A simple example can be found in the context of holographic graph models introduced in \cite{Bao:2015bfa}, by similar arguments as used in e.g.~\cite{Hernandez-Cuenca:2023iqh}: the graph describing perfect tensor (given by a 4-ray star graph with unit weights) on $\x{A,B,C,D}$  completed by isolated vertices gives $-\xctI{B...:C...:D...}{A...}=-2$ and vanishes whenever any of the main arguments does not contain one of $\x{B,C,D}$.  Hence on this simple graph model, the full TF evaluates to a negative number.
}
on which $Q<0$, i.e.\ the inequality is not a valid HEI.

\item $R$-balance: \\
Recall that all higher HEIs are superbalanced \cite{He:2020xuo} or equivalently 2-balanced.  Given this and the various structural constraints on the HEIs, the reader might be curious whether any could in fact be $R$-balanced for $R\ge 3$ (cf.~\cref{fn:Rbalance}).  Empirically this is not the case for any known HEI, but TF-compatibility ensures this result universally:  In particular, $R$-balance for $R\ge 3$ requires that the I-basis expansion of the corresponding information quantity does not contain any singleton-tripartite information terms such as $\sI_3(\x{A}\! : \! \x{B}\! : \! \x{C})$.
On the other hand, any TF expression, when expanded in the I-basis, always has such singleton-tripartite information terms appearing with a positive coefficient; for example, the term $\xctI{X:Y:Z}{W}$ has all possible singleton tripartite terms of the form $\sI_3({\xs{A}{i}}\! : \! {\xs{B}{j}}\! : \! {\xs{C}{k}})$  where $\xs{A}{i} \in \x{X}$, $\xs{B}{j} \in \x{Y}$ and $\xs{C}{k} \in \x{Z}$.   Since these always appear with a fixed sign,  the $\sI_3$ basis terms cannot get canceled between multiple TF terms, and therefore must remain in any TF expression. A more immediate way to see this is to observe that  the total number of terms in STF is positive, so using the \cno\ classification, $i_3 = \sum_\ell c_\ell>0$.
\end{itemize} 

\paragraph{Further repackaging:}
Given the utility of the TF packaging for revealing general structural properties of HEIs, one may naturally wonder if we can uncover an even more useful repackaging.  
As remarked above, one of the key missing ingredients is the TF-compatibility preservation under purifications.  
While any purification of any superbalanced HEI remains a superbalanced HEI, this is not the case for a general TF expression.  More specifically, while superbalance is retained, positivity is not, due to \cref{eq:CI3purif}.\footnote{\, 
    Hence  proving that one representative of an HEI is TF-compatible does not a priori mean that every element of its permutation-purification orbit is TF-compatible. 
    (For the $\nN=6$ HEIs of \cite{Hernandez-Cuenca:2023iqh}, the authors had checked that the purifications do admit a TF, albeit not necessarily of the same size.)
    For the \qtor\ and \qproj\ families, our TF expressions and the symmetries of these inequalities show that they are TF-compatible for any choice of purifier.
    More generally, the symmetries leaving an information quantity $Q$ unchanged allow us to partition the $\nN+1$ parties (including the purifier) into classes based on the stabilizer group.
    It then suffices to prove that $Q$ is TF-compatible for one choice of purifier in each class, since choices within the same class differ only by relabeling the parties. For the \qproj\ family, all parties belong to one class. For the \qtor\ family, we have two classes, i.e., it is enough to consider one A region and one B region as the purifier.
}
Nevertheless, if (as our evidence indicates) \cref{conj:HEIsTF} indeed holds, so that any superbalanced HEI is TF-compatible, it must be true that its purification is likewise TF-compatible -- which makes the tripartite forms for HEIs special.  In other words, the fact that this is not manifest from the TF structure suggests that individual TF terms are not the optimal building blocks for HEIs.  Nevertheless, since they already guarantee superbalance and I-basis sign alternation, any sums built out of them retain these desirable features.  This suggests that we should seek building blocks which are themselves TF-compatible, but composed of multiple $-I_3$ terms so as to \emph{manifestly} preserve TF-compatibility under purifications.  

Of course, the individual HEIs themselves do have this property, so one can try to build higher HEIs out of lower ones.\footnote{\, 
    Related ideas have been considered by various authors over the years.  For example, apart from the infinite families of \cite{Czech:2023xed} examined above, in \cite{Czech:2022fzb}, the authors uncovered a new $\nN=7$ HEI (which is in fact $\QT{5,3}$) by ``oxidizing'' lower HEIs. More closely-related to our present explorations, \cite{Hernandez-Cuenca:2023iqh} observed families of HEIs whose tripartite forms successively augment each other (cf.\ \cite[eq.(5.5)]{Hernandez-Cuenca:2023iqh}), as well as other families which appear to be composed of piecewise fine-grainings.
}
But since simply  summing up primitive HEIs gives a redundant HEI instead of a primitive one, they would not correspond to additive building blocks. Moreover, even if we formulate a prescription for composing HEIs non-additively, it may not be clear that this would generate all of them.  Instead, one can ask what are the smallest possible TF-compatible building blocks whose purifications remain TF-compatible.  If the answer only recovers the HEIs and the method is efficient, then we will have discovered a new way of generating the HEIs. If on the other hand we can find smaller building blocks, then we will have found a potentially superior repackaging.  We leave the explorations of this interesting question to future work.

\acknowledgments
We would like to thank 
Guglielmo Grimaldi,
Matt Headrick,
Sergio Hern\'andez-Cuenca,
Mukund Rangamani,
and
Max Rota
for useful discussions.
Y.L.\ would like to thank Di Pan and Jianming Zheng for their support during the summer in Davis, and the hospitality of QMAP at University of California, Davis, during early stages of this work.
V.H.\ was supported in part by the U.S. Department of Energy through award DE-SC0009999 and by funds from the University of California. 
Part of this work was performed at Aspen Center for Physics, which is supported by National Science Foundation grant PHY-2210452.
We used ChatGPT and Codex for help with the figures and the TF diagrammatic representation macro.

\appendix
\newpage

\section{Technical details}
\label{app:proof}

In this appendix, we collect the proofs relegated from the main text, in particular \cref{lmm:rela} in \cref{ss:lem2pf} and \cref{thm:projct} in \cref{ss:thm2pf}.  For completeness, we also list two alternate \ctform s\ of \qcyc\ inequalities in \cref{ss:qtorctform}.

\subsection{Proof of \texorpdfstring{\cref{lmm:rela}}{lemma 2}}
\label{ss:lem2pf}

To prove \cref{lmm:rela}, it will turn out convenient to recast the expressions back into the S basis.   This is because the expression for the difference between $\QT{m,n}$ and $\QT{m,1}$ is particularly simple in the S basis:  Using \cref{eq:Qtoric}, we see that the composite $A$-entropy in the last term cancels, and at $n=1$ the $B$ terms trivialize, leaving 
\begin{equation}\label{eq:QTdiff}
    \QT{m,1}-\QT{m,n} =
    \sum_{i = 1}^m \left( S_{A_i^+} - S_{A_i^-}\right)
    - \sum_{i=1}^m \sum_{k=1}^n \left( S_{A_i^+ B_k^-}-S_{A_i^- B_k^-} \right)  \ .
\end{equation}
To compare this to the TF expression, we start by considering the summand in \cref{eq:Ttoc}, in particular the term 
$\fctI{\xs{A}{i}}{\xs{B}{\nu+k}}{\xm{B}{k}{\nu-k+1}}{\xm{A}{i+1}{\mu}\xm{B}{\nu+1}{k-1}} $.
Structurally, this takes the abstract form $\xctI{X:Y:Z}{UV}$, which can be recast by iterating \cref{eq:CI3fromI3}, i.e.\
$\xctI{X:Y:Z}{UV} = \xctI{XU:Y:Z}{V}-\xctI{U:Y:Z}{V} =\xtI{XU:YV:Z}-\xtI{XU:V:Z} -\xtI{U:YV:Z}+\xtI{U:V:Z}$.  Rendered in the S basis this becomes
$
 - \x{XUYV}  + \x{XUYVZ}
 + \x{XUV}   - \x{XUVZ}
 + \x{UYV}  - \x{UYVZ}
 - \x{UV}   + \x{UVZ}
$
where we can simplify the adjoining terms $\x{XU}=\xm{A}{i}{\mu+1}=\xtp{A}{i}$, $\x{ZV}=\xm{B}{k}{\nu}=\xtm{B}{k}$, $\x{ZVY}=\xm{B}{k}{\nu+1}=\xtp{B}{k}$, and $\x{VY}=\xm{B}{\nu+1}{k}$, obtaining
\begin{equation}\label{eq:ExtraC}
    \begin{split}
         &\fctI{\xs{A}{i}}{\xs{B}{\nu+k}}{\xm{B}{k}{\nu-k+1}}{\xm{A}{i+1}{\mu}\xm{B}{\nu+1}{k-1}} \\
         = & +\fS{\xm{A}{i}{\mu+1}\xm{B}{k}{\nu+1}} +\fS{\xm{A}{i+1}{\mu}\xm{B}{k}{\nu}}+ \fS{\xm{A}{i+1}{\mu}\xm{B}{\nu+1}{k}}+\fS{\xm{A}{i}{\mu+1}\xm{B}{\nu+1}{k-1}}\\
        & - \fS{\xm{A}{i+1}{\mu}\xm{B}{k}{\nu+1}}-\fS{\xm{A}{i}{\mu+1}\xm{B}{k}{\nu}} - \fS{\xm{A}{i}{\mu+1}\xm{B}{\nu+1}{k}}-\fS{\xm{A}{i+1}{\mu}\xm{B}{\nu+1}{k-1}}\\
        = & +\fS{\xtp{A}{i}\xtp{B}{k}} +\fS{\xtm{A}{i+1}\xtm{B}{k}}+ \fS{\xtm{A}{i+1}\xm{B}{\nu+1}{k}}+\fS{\xtp{A}{i}\xm{B}{\nu+1}{k-1}}\\
        & -\fS{\xtm{A}{i+1}\xtp{B}{k}} -\fS{\xtp{A}{i}\xtm{B}{k}} - \fS{\xtp{A}{i}\xm{B}{\nu+1}{k}}-\fS{\xtm{A}{i+1}\xm{B}{\nu+1}{k-1}} \ .
    \end{split}
\end{equation}
After summing over $k$, most of the terms between the 4th and 7th expressions cancel, as do those between the 8th and 3rd expressions: 
\begin{equation}\label{eq:equS1}
    \begin{split}
        \sum_{k = 1}^{\nu} \left[ \fS{\xtp{A}{i}\xm{B}{\nu+1}{k-1}} - \fS{\xtp{A}{i}\xm{B}{\nu+1}{k}} \right] = - \fS{\xtp{A}{i}\xtm{B}{\nu+1}} + \fS{\xtp{A}{i}}
    \end{split}
\end{equation}
\begin{equation}\label{eq:equS2}
    \sum_{k = 1}^{\nu} \left[ -\fS{\xtm{A}{i+1}\xm{B}{\nu+1}{k-1}} + \fS{\xtm{A}{i+1}\xm{B}{\nu+1}{k}} \right]= + \fS{\xtm{A}{i+1}\xtm{B}{\nu+1}} - \fS{\xtm{A}{i+1}}
\end{equation}
To convert the remaining terms in \cref{eq:ExtraC} involving $\xtp{B}{k}$ to those involving $\xtm{B}{k}$, we use purity, in particular
$\fS{\xtp{A}{i}\xtp{B}{k}} =\fS{\xtm{A}{i+\mu+1}\xtm{B}{k+\nu+1}}$ and 
$\fS{\xtm{A}{i+1}\xtp{B}{k}} = \fS{\xtp{A}{i+\mu+1}\xtm{B}{k+\nu+1}}$. 
Now, notice that in summing over $i=1,\ldots,m$, we are summing one full cycle on the $A$ regions, which means that we can shift the $i$ index at will under this sum; for example, we can shift all $A$ subscripts to start at $i$.   On the other hand, the $k$ sum is \emph{not} over the full cycle, since $k=1,\ldots, \nu$.  However the three contributions from the positive terms combine to a full $B$-cycle,
\begin{equation}\label{eq:}
    \begin{split}
& \sum_{i = 1}^{m} 
    \left(
        \sum_{k = 1}^{\nu} 
        \left[
            \fS{\xtm{A}{i+\mu+1}\xtm{B}{k+\nu+1}}
            +\fS{\xtm{A}{i+1}\xtm{B}{k}}
        \right] 
        + \fS{\xtm{A}{i+1}\xtm{B}{\nu+1}} 
    \right)
\\
& =
 \sum_{i = 1}^{m} 
    \left(
        \sum_{k = 1}^{\nu} 
        \left[
            \fS{\xtm{A}{i}\xtm{B}{k+\nu+1}}
            +\fS{\xtm{A}{i}\xtm{B}{k}}
        \right] 
        + \fS{\xtm{A}{i}\xtm{B}{\nu+1}} 
    \right)
\\
& =
\sum_{i = 1}^{m} \sum_{k = 1}^{n} \fS{\xtm{A}{i}\xtm{B}{k}}  \ ,
    \end{split}
\end{equation}
and similarly for the three negative contributions.
Putting everything together and using the simplification 
\cref{eq:QTdiff}, we finally obtain
\begin{equation}\label{eq:toirccyc}
    \begin{split}
        & \sum_{i = 1}^{m}\sum_{k = 1}^{\nu} \fctI{\xs{A}{i}}{\xs{B}{\nu+k}}{\xm{B}{k}{\nu-k+1}}{\xm{A}{i+1}{\mu}\xm{B}{\nu+1}{k-1}}\\
        = & - \sum_{i=1}^m \sum_{k=1}^n \left( S_{A_i^+ B_k^-}-S_{A_i^- B_k^-} \right) + \sum_{i = 1}^m \left( S_{A_i^+} - S_{A_i^-}\right) \\
        = &  \QT{m,1} - \QT{m,n} 
    \end{split}
\end{equation}
This proves \cref{lmm:rela}.
\qed

\subsection{Proof of \texorpdfstring{\cref{thm:projct}}{theorem 2}}
\label{ss:thm2pf}

We first recall that we adopt the convention for the \qproj\ inequalities given in \cref{eq:Qproj}. As in the proof of \cref{lmm:rela}, we begin by expressing the proposed \ctform\ in the S basis:
\begin{equation}\label{eq:Sproj}
\begin{split}
    \sum_{i = 1}^{m-1} \sum_{j = i}^{m-1} &- \fctI{\xs{A}{i}}{\xs{B}{j}}{\xm{A}{j+1}{m-j} }{\xm{A}{1}{i-1}\xm{B}{i}{j-i}}\\
     =  \sum_{i = 1}^{m-1} \sum_{j = i}^{m-1}\ \Big[ & \fS{\xm{A}{1}{i-1}\xm{B}{i}{j-i}} -\fS{\xm{A}{1}{i-1}\xm{B}{i}{j-i+1}}\\
     + &\fS{\xm{A}{1}{i}\xm{B}{i}{j-i+1}} - \fS{\xm{A}{1}{i}\xm{B}{i}{j-i}}\\
     - & \fS{\xm{A}{j+1}{m+i-j}\xm{B}{i}{j-i+1}} - \fS{\xm{A}{j+1}{m+i-j-1}\xm{B}{i}{j-i}}\\
     + & \fS{\xm{A}{j+1}{m+i-j}\xm{B}{i}{j-i}} + \fS{\xm{A}{j+1}{m+i-j-1}\xm{B}{i}{j-i+1}} \Big]
\end{split}
\end{equation}

Summing the first two lines on the right-hand side of \cref{eq:Sproj} over the index $j$ yields:
\begin{equation}\label{eq:sumj}
    \begin{split}
        & \sum_{i = 1}^{m-1} \sum_{j = i}^{m-1}\  \left[ \fS{\xm{A}{1}{i-1}\xm{B}{i}{j-i}} -\fS{\xm{A}{1}{i-1}\xm{B}{i}{j-i+1}}+\fS{\xm{A}{1}{i}\xm{B}{i}{j-i+1}} - \fS{\xm{A}{1}{i}\xm{B}{i}{j-i}} \right]\\
         = & \sum_{i = 1}^{m-1}\  \left[ \fS{\xm{A}{1}{i-1}} - \fS{\xm{A}{1}{i-1}\xm{B}{i}{m-i}}+ \fS{\xm{A}{1}{i}\xm{B}{i}{m-i}} - \fS{\xm{A}{1}{i}} \right] \\
          = & \sum_{i = 1}^{m-1}\  \left[ \fS{\xm{A}{1}{i}\xm{B}{i}{m-i}} - \fS{\xm{A}{1}{i-1}\xm{B}{i}{m-i}} \right]  - \fS{\xm{A}{1}{m-1}} 
    \end{split}
\end{equation}
Substituting this result into \cref{eq:Sproj}, we group the terms according to their signs and denote the resulting contributions by $Q_{(m)}^+$ and $Q_{(m)}^-$. Explicitly,
\begin{equation}\label{eq:plus}
    \begin{split}
        Q_{(m)}^+ =& \sum_{i = 1}^{m-1} \sum_{j = i}^{m-1} \left[ \fS{\xm{A}{j+1}{m+i-j}\xm{B}{i}{j-i}} +\fS{\xm{A}{j+1}{m+i-j-1}\xm{B}{i}{j-i+1}} \right]\\
        & + \sum_{i = 1}^{m-1}\ \fS{\xm{A}{1}{i}\xm{B}{i}{m-i}}
    \end{split}
\end{equation}    
\begin{equation}\label{eq:minus}
    \begin{split}
        Q_{(m)}^- =& \sum_{i = 1}^{m-1} \sum_{j = i}^{m-1}\  \left[ - \fS{\xm{A}{j+1}{m+i-j}\xm{B}{i}{j-i+1}} - \fS{\xm{A}{j+1}{m+i-j-1}\xm{B}{i}{j-i}} \right]\\
        & + \sum_{i = 1}^{m-1}\ \left[- \fS{\xm{A}{1}{i-1}\xm{B}{i}{m-i}} \right] - \fS{\xm{A}{1}{m-1}} \\
         = & -\sum_{i = 1}^{m}\sum_{j = 1}^{m} \fS{\xm{A}{i}{j-1}\xm{B}{i+j-1}{m-j}}
    \end{split}
\end{equation}
The final line of \cref{eq:minus} follows from relabeling the summation indices and regrouping the terms. It coincides precisely with the minus sign terms in \cref{eq:Qproj}. It therefore remains to simplify the $Q_{(m)}^+$ terms.

Combining the first term in the double sum in $Q_{(m)}^+$ with its final sum, we obtain
\begin{equation}\label{eq:groupfl}
    \begin{split}
        &  \sum_{i = 1}^{m-1} \sum_{j = i}^{m-1} \left[ \fS{\xm{A}{j+1}{m+i-j}\xm{B}{i}{j-i}} \right] +  \sum_{i = 1}^{m-1}\ \fS{\xm{A}{1}{i}\xm{B}{i}{m-i}} \\
        = & \sum_{i = 2}^{m}\sum_{j = m-i+2}^{m} \left[ \fS{\xm{A}{i}{j}\xm{B}{i+j-1}{m-j}}\right] + \sum_{j = 1}^{m-1}\ \fS{\xm{A}{1}{j}\xm{B}{j}{m-j}} \\
        = & \frac{1}{2} \sum_{i = 1}^{m}\sum_{j = 1}^{m} \left[ \fS{\xm{A}{i}{j}\xm{B}{i+j-1}{m-j}} \right] + \frac{1}{2} m \cdot \fS{\xs{A}{1}\cdots \xs{A}{m}} - \fS{\xs{A}{1}\cdots \xs{A}{m}} 
    \end{split}
\end{equation}
where the factor of $\frac{1}{2}$ takes care of the extended summation limits.
The remaining term can similarly be simplified as
\begin{equation}\label{eq:groupm}
    \begin{split}
        \sum_{i = 1}^{m-1} \sum_{j = i}^{m-1} \left[ \fS{\xm{A}{j+1}{m+i-j-1}\xm{B}{i}{j-i+1}} \right]  = \frac{1}{2} \sum_{i = 1}^{m}\sum_{j = 1}^{m-1} \fS{\xm{A}{i}{j}\xm{B}{i+j}{m-j}}
    \end{split}
\end{equation}
Combining these results gives
\begin{equation}\label{eq:Qplus}
    \begin{split}
        Q_{(m)}^+ & =  \frac{1}{2} \sum_{i = 1}^{m}\sum_{j = 1}^{m} \left[ \fS{\xm{A}{i}{j}\xm{B}{i+j-1}{m-j}} \right] + \frac{1}{2} \sum_{i = 1}^{m}\sum_{j = 1}^{m-1} \fS{\xm{A}{i}{j}\xm{B}{i+j}{m-j}} + \frac{1}{2} m \cdot \fS{\xs{A}{1}\cdots \xs{A}{m}} - \fS{\xs{A}{1}\cdots \xs{A}{m}} \\
        & = \frac{1}{2} \sum_{i = 1}^{m}\sum_{j= 1}^{m} \left[ \fS{\xm{A}{i}{j}\xm{B}{i+j-1}{m-j}} + \fS{\xm{A}{i}{j}\xm{B}{i+j}{m-j}}\right] - \fS{\xs{A}{1}\cdots \xs{A}{m}}
    \end{split}
\end{equation}
Finally, adding $Q_{(m)}^+$ and $Q_{(m)}^-$, we obtain
\begin{equation}\label{eq:RPtri}
    \begin{split}
        \QP{m} & =  Q_{(m)}^+ +  Q_{(m)}^-\\
           & = \frac{1}{2} \sum_{i = 1}^{m}\sum_{j= 1}^{m} \left[ \fS{\xm{A}{i}{j}\xm{B}{i+j-1}{m-j}} + \fS{\xm{A}{i}{j}\xm{B}{i+j}{m-j}}\right] -\sum_{i = 1}^{m}\sum_{j = 1}^{m} \fS{\xm{A}{i}{j-1}\xm{B}{i+j-1}{m-j}} - \fS{\xs{A}{1}\cdots \xs{A}{m}}
    \end{split}
\end{equation}
This expression agrees exactly with \cref{eq:Qproj}. Therefore, \cref{eq:projct} gives the \ctform\ of the \qproj\ inequalities.
\qed

\subsection{Alternate \ctform s\ of \qcyc\ inequalities}
\label{ss:qtorctform}

As we have seen above, non-trivial TF-compatible expressions admit multiple \ctform\ representations, which may expose different structural properties.   Since the \qcyc\ inequalities were the trickiest ones to convert into TF, here we collect two alternative TF representations of $\QT{m,1}$. These follow from similar manipulations as illustrated above; we leave their proofs as an exercise for the reader.

The expression in \cref{eq:Qm1} for the \ctform\ of the \qcyc\ inequality consists of two sets of terms, but in fact we can combine them into a single expression. To that end, we first formally rewrite it in a piecewise form:
\begin{equation}\label{eq:qtm1a}
    \QT{m,1} = -\sum_{a=1}^\mu \sum_{j=1}^\mu 
      \fctI{\x{A}^{*}}{\xs{A}{-a}}{\xs{A}{j-a}}{\xm{A}{\mu+1}{\mu-a} \xm{A}{j-a+1}{\mu-j} }, 
      \quad \text{where } 
      \x{A}^{*} = 
      \begin{cases}
        \xm{A}{\mu-a+1}{a}, &  j \geq a, \\
        \xm{A}{\mu-a+1}{j}, &  j < a
      \end{cases}
\end{equation}
where we used the conventions of \cref{eq:toriconvention} and the cyclicity of indices defined below \cref{eq:Akdef}.
This form can be further simplified by using a minimization over the indices:
\begin{equation}\label{eq:qtm1b}
    \QT{m,1} = -\sum_{i=1}^\mu \sum_{j=1}^\mu 
  \fctI{\xm{A}{\mu-i+1}{\min \{i,j\}}}{\xs{A}{-i}}{\xs{A}{j-i}}{\xm{A}{\mu+1}{\mu-i}\xm{A}{j-i+1}{\mu-j}} .
\end{equation}

Another TF expression, which we used for deriving one possible singleton tripartite form in \cref{s:toric}, is
\begin{equation}\label{eq:Qm12a}
    \begin{split}
        \QT{m,1} = & - \sum_{k = 1}^{\mu} \sum_{i = 1}^{\mu-k+1} \fctI{\xm{A}{k}{i}}{\xs{A}{\mu+k}}{\xs{A}{\mu+k+i}}{\xm{A}{\mu+1}{k-1}\xm{A}{\mu+k+i+1}{\mu-i}}\\
        & - \sum_{i = 1}^{\mu-1} \fctI{\xm{A}{1}{i}}{\xs{A}{\mu+i+1}}{\xm{A}{i+1}{\mu-i}}{\xm{A}{\mu+1}{i}} \ .
    \end{split}
    \end{equation}
To see the difference between this form and the one used in the main text, consider the diagrammatic representation, say for $m=11$ (so as to extend \cref{tab:QTgraphicalrep}) and rotated for compactness.  The first represents  \cref{eq:Qm1}  (which is equivalent to \cref{eq:torct} evaluated at $n=1$) and second one is \cref{eq:Qm12a}.
\begin{equation*}
\QT{11,1} = 
    \vTFtabHor[np,cell size=0.26,box size=0.13]{11}{{F, A, GHIJK, BCDE}, {F, B, HIJK, CDEG}, {G, B, HIJK, CDE}, {F, C, IJK, DEGH}, {G, C, IJK, DEH}, {H, C, IJK, DE}, {F, D, JK, EGHI}, {G, D, JK, EHI}, {H, D, JK, EI}, {I, D, JK, E}, {F, E, K, GHIJ}, {G, E, K, HIJ}, {H, E, K, IJ}, {I, E, K, J}, {J, E, K}, {G, BCDE, A, HIJK}, {H, CDE, AB, IJK}, {I, DE, ABC, JK}, {J, E, ABCD, K}}
=
    \vTFtabHor[np,cell size=0.26,box size=0.13]{11}{{A, F, G, HIJK}, {AB, F, H, IJK}, {ABC, F, I, JK}, {ABCD, F, J, K}, {ABCDE, F, K}, {B, G, H, AFIJK}, {BC, G, I, AFJK}, {BCD, G, J, AFK}, {BCDE, G, K, AF}, {C, H, I, ABFGJK}, {CD, H, J, ABFGK}, {CDE, H, K, ABFG}, {D, I, J, ABCFGHK}, {DE, I, K, ABCFGH}, {E, J, K, ABCDFGHI}, {A, G, BCDE, F}, {AB, H, CDE, FG}, {ABC, I, DE, FGH}, {ABCD, J, E, FGHI}}
\end{equation*}
However, both are the same length, so neither is a priori superior.

\newpage
\bibliography{references}
\bibliographystyle{JHEP}

\end{document}